\documentclass[prd,aps,twocolumn,showpacs,superscriptaddress]{revtex4-1}

\usepackage{array}
\usepackage{amsmath}
\usepackage{graphicx}
\usepackage{amssymb}

\usepackage{color}
\usepackage[colorlinks,citecolor=blue]{hyperref}
\usepackage{subfigure}
\usepackage{bm}
\usepackage[colorinlistoftodos]{todonotes}
\usepackage{mathtools}

\providecommand{\aap}{Astron.\ Astrophys.}
\providecommand{\apjl}{Astrophys.\ J.}
\providecommand{\apjs}{Astrophys.\ J. Suppl.}
\providecommand{\physrep}{Phys.\ Rep.}
\providecommand{\pasa}{Publ.\ Astron.\ Soc.\ Austr.}
\providecommand{\epja}{Eur.\ Phys.\ J.\ A}
\providecommand{\mnras}{Month.\ Not.\ Roy.\ Astron.\ Soc.}

\makeatletter

\begin{document}

\title{Muonization of supernova matter}

\author{Tobias Fischer}\email{tobias.fischer@uwr.edu.pl}
\affiliation{Institute of Theoretical Physics, University of Wroc{\l}aw, 50-204 Wroc{\l}aw, Poland}

\author{Gang Guo}
\affiliation{Institute of Physics, Academia Sinica, Taipei, 11529, Taiwan}

\author{Gabriel Mart{\'i}nez-Pinedo}
\affiliation{GSI Helmholtzzentrum f\"ur Schwerioneneforschung, 64291 Darmstadt, Germany}
\affiliation{Institut f{\"u}r Kernphysik, Technische Universit{\"a}t Darmstadt, 64289 Darmstadt, Germany}
\affiliation{Helmholtz Forschungsakademie Hessen f{\"u}r FAIR, GSI Helmholtzzentrum f{\"u}r Schwerionenforschung, Planckstra{\ss}e~1,  64291 Darmstadt, Germany}

\author{Matthias Liebend{\"o}rfer}
\affiliation{Department of Physics, University of Basel, 4056 Basel, Switzerland}

\author{Anthony Mezzacappa}
\affiliation{Department of Physics and Astronomy, University of Tennessee, Knoxville, TN 37996-1200, USA}

\begin{abstract}

\bigskip

The present article investigates the impact of muons on core-collapse supernovae, 
with particular focus on the early muon neutrino emission. While the presence of 
muons is well understood in the context of  neutron stars, until the recent study
by Bollig~{\em et al.}~[Phys.~Rev.~Lett.~\textbf{119},~242702~(2017)] the role of muons
in core-collapse supernovae had been neglected---electrons and neutrinos were 
the only leptons considered. In their study, Bollig~{\em et~al.} disentangled the muon 
and tau neutrinos and  antineutrinos and included a variety of muonic weak 
reactions, all of which the present paper follows closely. Only then does it 
becomes possible to quantify the appearance of muons shortly before stellar 
core bounce and how the post-bounce prompt neutrino emission is modified. \\ \\
\bigskip
DOI: 10.1103/PhysRevD.102.123001\\
\end{abstract}

\received{31 august 2020}\accepted{21 October 2020}\published{1 December 2020}


\maketitle

\section{Introduction}
\label{sec:intro}
A core-collapse supernova (SN) determines the final fate of all stars more massive 
than  about 8~M$_\odot$. The associated stellar core collapse is triggered due to 
deleptonization by nuclear electron capture in the core and the subsequent escape 
of the electron neutrinos produced, lowering the degenerate electron density responsible 
for supporting the core against gravity, and to the photo-disintegration of heavy 
nuclei in the core, sapping thermal, pressure-producing energy as well. The collapse 
halts when the central density exceeds normal nuclear density. The repulsive 
short-range nuclear force reverses the collapse, and the stellar core rebounds. 
An expanding shock wave forms, which stalls when crossing the neutrinospheres, 
the surfaces of last scattering for the neutrinos produced and trapped during 
core collapse. A large number of electron captures on the newly liberated 
protons from the dissociation of nuclei by the shock releases the deleptonization 
burst after the shock passes the neutrinospheres. 
This happens on a timescale of about 5--20~ms after core 
bounce~\cite{Janka:2007,Janka:2012}. The central compact object comprising a 
cold, un-shocked core and a hot, shocked mantle is the proto-neutron star (PNS). 
The so-called SN problem poses the question: How is the stalled bounce shock 
revived? Several scenarios have been proposed: the neutrino heating~\cite{Bethe:1985ux}, 
magneto-rotational~\cite{LeBlanc:1970kg}, and acoustic~\cite{Burrows:2005dv} 
mechanisms, as well as a mechanism associated with a high-density phase 
transition in the core~\cite{Sagert:2008ka,Fischer:2018,Fischer:2020b}. 
Studies of the multi-physics, multi-scale core-collapse SN phenomenology 
require large-scale computer models, which are based on neutrino radiation-
hydrodynamics. For a recent review about the various scales and conditions 
of relevance, as well as the SN equation of state (EOS), 
cf.~Ref.~\cite{Mueller2016,Fischer:2017}.

During a core-collapse SN, the neutrinos propagate through regions
that are diffusive, semitransparent, and transparent (where the
neutrinos simply stream freely). Thus, the neutrinos are not
fluid-like everywhere, and a full Boltzmann kinetic treatment of
neutrino transport is ultimately necessary. This has been achieved in
the context of general relativistic models in spherical symmetry
\cite{Yamada:etal:1999,Liebendoerfer:2001b} and in axisymmetry
\cite{Mueller:2020}, as well as non-relativistic and
relativistic axisymmetric models with Newtonian gravity
\cite{OtBuDe08,NaIwFu18}. While pioneering and already advancing
with respect to treating separately $\nu_{\mu/\tau}$ and 
$\bar\nu_{\mu/\tau}$, all of these studies suffer from a draw 
back: they assume equal distributions of $\mu$- and
$\tau$-neutrinos and antineutrinos.

This simplification can only be
justified in the absence of muons. However, it is well known for cold
neutron stars, where due to the condition of $\beta$-equilibrium muons
and electrons have equal chemical potentials ($\mu_\mu=\mu_e$).
Hence, when $\mu_e>m_\mu\simeq106$~MeV, the
muon fraction can be as large as $Y_\mu\simeq0.02-0.05$ (depending
on the nuclear EOS) above a rest-mass density of about
half the saturation density ($2.5\times 10^{14}$~g~cm$^{-3}$). The
presence of muons has important consequences for the long-term cooling
of neutron stars; e.g., it modifies the direct-Urca
threshold~\cite{Glendenning_book}. Muons have to be
produced at some point during the evolution of the PNS from a 
hot lepton rich object to the cold $\beta$-equilibrium
object discussed above. However, this aspect has only been recently 
studied in Ref.~\cite{Bollig:2017}, including the possibility of 
muons decaying to axions \cite{Bollig:2020}.
Muons can be produced in non-negligible abundances 
during a core-collapse SN. The present article extends this study 
to consider the muonization of SN matter shortly before core bounce and 
discusses the impact of the presence of muons on the neutrino emission 
up to shortly after core bounce.
Therefore, the Boltzmann neutrino transport scheme is extended to treat
$\mu$- and $\tau$-neutrinos and antineutrinos separately, include
an extended set of weak processes with muons in the collision integral
on the right-hand side of the Boltzmann equation, and add the muon
abundance as an additional independent degree of freedom in the
radiation-hydrodynamics scheme.

The present article is organized as follows. In Sec.~\ref{sec:SNmodel},
the SN model is briefly reviewed, with special emphasis placed on the 
updates to the neutrino transport scheme. Sec.~\ref{sec:sim} discusses 
our SN simulation results in close proximity of stellar core bounce, 
with a focus on the muonization of SN matter and on the enhanced 
muon-neutrino luminosity.
In Sec.~\ref{sec:convection}, we consider the possibility for convection 
to occur due to the presence of what will now be an additional lepton 
number gradient. The manuscript closes with a summary in 
Sec.~\ref{sec:summary}.

\section{Core-collapse supernova model}
\label{sec:SNmodel}
The core-collapse SN model employed in this study, {\tt AGILE-BOLTZTRAN}, is based on general relativistic neutrino radiation hydrodynamics in spherical symmetry~\cite{Mezzacappa:1993gn,Mezzacappa:1993gm,Mezzacappa:1993gx}, in comoving coordinates~\cite{Liebendoerfer:2001a,Liebendoerfer:2004} with a Lagrangian mesh featuring an adaptive mesh refinement method~\cite{Liebendoerfer:2002}. In the present study 207 radial mass zones are used. A recent global-comparison core-collapse SN study in spherical symmetry, including {\tt AGILE-BOLTZTRAN}, can be found in Ref.~\cite{OConnor:2018}.

\subsection{Equation of state}
\label{sec:SNmodel_EOS}
{\tt AGILE-BOLTZTRAN} has a flexible EOS module treating separately
the nuclear part~\cite{Hempel:2012} and the
electron/positron/photon/Coulomb EOS; the latter is collectively
denoted as EPEOS~\cite{Timmes:1999}. In addition to the temperature,
$T$, and restmass density, $\rho$, the EOS depends also on the nuclear
composition with mass fractions $X_i$, atomic mass $A_i$ and charge
$Z_i$. The latter determines the charge fraction of the nuclei, which 
balances the combined charge fractions of electrons, $Y_e$ and muons, 
$Y_\mu$. Here, the nuclear EOS of Ref.~\cite{Hempel:2009mc} is 
employed. It is based on the modified nuclear statistical
equilibrium approach for several 1000 nuclear species and the density-
dependent relativistic mean-field model DD2~\cite{Typel:2009sy}
for the unbound nucleons.

In the present study, a muon EOS is implemented in
{\tt AGILE-BOLTZTRAN}. Therefore, the following muon EOS quantities
are tabulated: particle density $n_\mu=Y_\mu n_{\rm B}$, internal energy
density $e_{\mu^\pm}$, pressure $P_{\mu^\pm}$ and entropy per particle
$s_{\mu^\pm}$, as a function of the muon chemical potential ranging
$\mu_\mu=0,\ldots,500$~MeV for a large range of temperatures from
$T=0,\ldots,200$~MeV. The Fermi integrals are performed numerically
with a 64-point Gauss-quadrature. This ensures thermodynamic
consistency. Since muons are massive leptons, their restmass cannot be
neglected, and the relativistic dispersion relation must be employed,
$E=\sqrt{p^2+m_\mu^2}$, unlike electrons/positrons, which are
ultrarelativistic ($E\simeq p$). In the SN simulations, where the muon
abundance becomes the degree of freedom for the muon EOS, in addition
to temperature and restmass density, a linear interpolation is used to
find the corresponding muonic thermodynamic state,
$\mu_\mu(T,\rho Y_\mu)$, $e_{\mu^\pm}(T,\rho Y_\mu)$,
$P_{\mu^\pm}(T,\rho Y_\mu)$ and $s_{\mu^\pm}(T,\rho Y_\mu)$,
respectively. These quantities, except $\mu_\mu$, are then added to
the corresponding quantities for baryons (B) and EPEOS, in order to
obtain the total quantities,
\begin{eqnarray}
e_{\rm tot} &=& e_{\rm B}(T,\rho,Y_p) + e_{\rm EPEOS}(T,\rho Y_e,\{X_i,A_i,Z_i\}) \nonumber \\
&&\; +  e_{\mu^\pm}(T,\rho Y_\mu)~, \label{eq:e}\\
\nonumber \\
P &=& P_{\rm B}(T,\rho,Y_p) + P_{\rm EPEOS}(T,\rho Y_e,\{X_i,A_u,Z_i\})  \nonumber \\
&&\; +  P_{\mu^\pm}(T,\rho Y_\mu)~,  \label{eq:P}\\
\nonumber \\
s &=& s_{\rm B}(T,\rho,Y_p) + s_{\rm EPEOS}(T,\rho Y_e,\{X_i,A_i,Z_i\}) \nonumber \\
&&\; +  s_{\mu^\pm}(T,\rho Y_\mu)~.\label{eq:s}
\end{eqnarray}
Note that the baryon EOS contributions depend on the hadronic charge
fraction via the charge-neutrality conditions, $Y_p=Y_e+Y_\mu$, where
electron and muon abundances are associated with their corresponding
net particle densities, such that $Y_e=Y_{e^-}-Y_{e^+}$ and 
$Y_\mu=Y_{\mu^-}-Y_{\mu^+}$.

\subsection{Boltzmann neutrino transport}
\label{sec:SNmodel_transport}
The neutrino transport scheme has to be extended in order to be able
to treat individually the distributions for all 3 flavors,
$\{f_{\nu_e},f_{\nu_\mu},f_{\nu_\tau}\}$ and their respected
antineutrinos $\{f_{\bar\nu_e},f_{\bar\nu_\mu},f_{\bar\nu_\tau}\}$. 
 {\tt BOLTZTRAN} employs an operator-split method to solve the
evolution equations for the neutrino distribution functions,
as described in detail in Ref.~\cite{Liebendoerfer:2004}
(steps 1.--3. outlined in sec.~3.5). 
Each implicitly finite differencing update of the transport
equation includes the update of the evolution of the temperature
and the electron/muon fraction due to weak interactions, as well
as corrections due to advection.
A Newton-Raphson scheme is implemented to solve the implicitly
finite differencing nonlinear equations.
Beginning with ($\nu_e,\bar\nu_e$), this procedure is repeated
for ($\nu_\mu,\bar\nu_\mu$) and ($\nu_\tau,\bar\nu_\tau$).
As was outlined already in Ref.~\cite{Liebendoerfer:2004}, this
cycling ensures that the neutrino distribution functions are in
accurate equilibrium with matter after obtaining the solution of the
transport equation and updating temperature and electron fraction
accordingly. However, it introduces a mismatch of the radial grid
of the new hydrodynamics variables due to the corrections of the
advection equation. 
Note further that for the weak processes,
$\nu_e+\bar\nu_e\leftrightarrows \nu_{\mu/\tau}+\bar\nu_{\mu/\tau}$,
where initial the final state neutrino distributions belong to
different species \cite{Buras:2005rp}, in the collision integral
of the Boltzmann transport equation we assume equilibrium
distributions as was outlined in Ref.~\cite{Fischer:2009}.
This approach is inadequate for the purely leptonic weak
processes involving muons, which will be introduced below, where
initial the final state neutrino distributions also belong to
different species. 
Here we implement for the final states the actual neutrino
distributions from the previous cycling, which introduces a slight
mismatch that we monitor carefully with an increased accuracy
required for the neutrino transport convergence.

Note further that {\tt BOLTZTRAN} employs the transport equation in
conservative form; i.e. with the specific neutrino distribution
function, $F_\nu:=f_\nu/\rho$~\cite{Mezzacappa:1993gn,Mezzacappa:1993gm}.
All neutrino species are discretized in terms of 6 momentum angles bins
$\cos\vartheta\in\{-1,+1\}$~\footnote{$\cos\vartheta$ is used here and
not $\mu=\cos\vartheta$ as in the standard literature, not to confuse 
with the muons} -- the angle between the radial motion and the 
momentum vector -- and 36 neutrino energy bins, $E_\nu\in\{0.5,300\}$~MeV 
following the setup of S.~Bruenn~\cite{Bruenn:1985en}. Appendix~\ref{sec:A} 
compares two SN simulations, both without muonic weak reactions, comparing 
the {\em traditional} Boltzmann transport scheme for 4 neutrino species
($\nu_e$, $\bar\nu_e$, $\nu_{\mu/\tau}$, $\bar\nu_{\mu/\tau}$) and 
the full 6 neutrino species transport ($\nu_e$, $\bar\nu_e$, $\nu_{\mu}$, 
$\bar\nu_{\mu}$, $\nu_{\tau}$, $\bar\nu_{\tau}$).
While the extension to 6-species Boltzmann neutrino transport is 
straight forward, the inclusion of weak interactions and the 
associated extensions of the collision-integral involving weak 
reactions with (anti)muons will be discussed in the next sections.
Further details are given in Appendix~\ref{sec:B}.

\subsection{Muonic weak processes}
In the following subsections all new weak reactions involving 
(anti)muons, which are being implemented in {\tt AGILE-BOLTZTRAN}, 
will be discussed. Table~\ref{tab:muonic-weak} lists all 
processes considered here. Further details, about the reaction 
rates are provided in Appendix~\ref{sec:B}, see also 
Ref.~\cite{Guo:2020}, and details about their implementation 
in {\tt AGILE-BOLTZTRAN} are provided in Appendix~\ref{sec:B} of the present paper.

\subsubsection{Charged-current absorption and emission}
For the emissivity
$\left(j_{\nu_\mu}(E_{\nu_\mu}),
  j_{\bar\nu_\mu}(E_{\bar\nu_\mu})\right)$ and absorptivity
$\left(\chi_{\nu_\mu}(E_{\nu_\mu}),
  \chi_{\bar\nu_\mu}(E_{\bar\nu_\mu})\right)$ for the muonic
charged-current (CC) reactions (1a) and (1b) in 
Table~\ref{tab:muonic-weak}, the fully inelastic and relativistic rates 
are employed. These rates were developed in Ref.~\cite{Guo:2020}, 
Section~III~A, which is based on the same treatment as for the 
electronic CC rates~\cite{Fischer:2020}. Ref.~\cite{Oertel:2020}
consider correlations at the level of the random-phase approximation
but neglect contributions from weak-magnetism which we take
self-consistently into account. Furthermore, the rate expressions of
Ref.~\cite{Guo:2020} consider pseudo-scalar interaction
contributions. 

\begin{table}[t!]
\caption{Set of muonic weak processes considered}
\label{tab:muonic-weak}
\begin{ruledtabular}
\begin{tabular}{lcc}
Label & Weak process & Abbreviation \\
 \hline
(1a) & $\nu_\mu + n \leftrightarrows p + \mu^-$ & CC \\
(1b) & $\bar\nu_\mu + p \leftrightarrows n + \mu^+$ & CC \\
& & \\
(2a) & $\nu_{\mu} + \mu^\pm \leftrightarrows \mu'^\pm + \nu'_{\mu}$ & NMS \\
(2b) & $\bar\nu_{\mu} + \mu^\pm \leftrightarrows \mu'^\pm + \bar\nu'_{\mu}$ & NMS \\
& & \\
(3a) & $\nu_{\mu} + e^- \leftrightarrows \mu^- + \nu_e$ & LFE \\
(3b) & $\bar\nu_{\mu} + e^+ \leftrightarrows \mu^+ + \bar\nu_e$ & LFE \\
(4a) & $\nu_{\mu} + \mu^+ \leftrightarrows e^+ + \nu_e$ & LFC \\
(4b) & $\bar\nu_{\mu} + \mu^- \leftrightarrows e^- + \bar\nu_e$ & LFC \\
\end{tabular}
\end{ruledtabular}
\end{table}
\begin{figure}[b!]
\centering
\includegraphics[width=1.0\columnwidth]{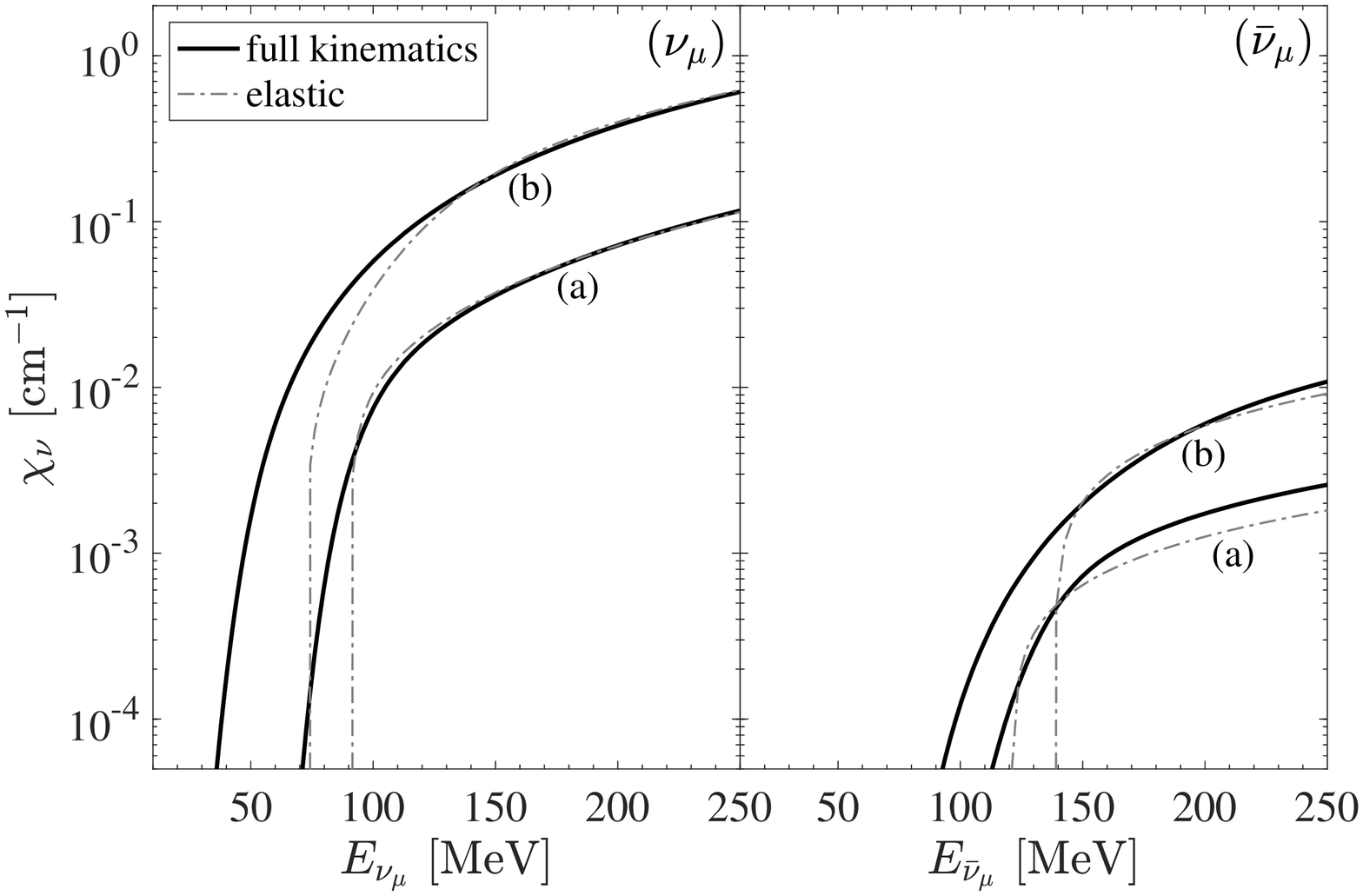}\caption{Neutrino ($\nu_\mu$)  and anti-neutrino ($\bar\nu_\mu$)
  opacity for the muonic charged-current reactions (1a) and (1b) in
  Table~\ref{tab:muonic-weak}, comparing the fully inelastic rates
  (Eq.~(33) in Ref.~\cite{Guo:2020}) (solid lines) and the elastic
  approximation \eqref{eq:cc_rate} (dashed lines), at two selected
  conditions referred to as (a) and (b) for which the corresponding
  thermodynamic state is given in Table~\ref{tab:eos}.} 
\label{fig:cc}
\end{figure}
\begin{table*}[t!]
\centering
\caption{Thermodynamic state for two selected conditions.}
\label{tab:eos}
\begin{ruledtabular}
\begin{tabular}{ccccccccccccccccccccccc}
 &&& $T$ &&& $\rho$ &&& $Y_e$ &&& $Y_\mu$  &&& $\mu_e$ &&& $\mu_\mu$ &&& $U_n-U_p$\footnote{$U_{n/p}$ are the neutron/proton single-particle potentials, which are given by the DD2 EOS} \\
 &&& $[$MeV$]$ &&& $[$g~cm$^{-3}]$ &&& &&& &&& $[$MeV$]$ &&& $[$MeV$]$ &&& $[$MeV$]$ \\
 \hline
(a) &&& 10 &&& $5\times 10^{13}$ &&& 0.2 &&& $10^{-4}$ &&& 108.1 &&& 51.7 &&& 13.9 \\
(b) &&& 25 &&& $2\times 10^{14}$ &&& 0.15 &&& 0.05 &&& 147.4 &&& 132.8 &&& 31.5 \\
\end{tabular}
\end{ruledtabular}
\end{table*}

Equations~(27)--(33) in Ref.~~\cite{Guo:2020}, as well as their Appendix~(B),
summarize the entire algebraic expressions. Since the transition amplitudes
-- the spin averaged and squared matrix elements -- are identical for 
electronic and muonic charged-current reactions, the only difference is 
the remaining phase space. Hence, the only replacements for the muonic 
charged-current rates are the different muon restmass and the muon Fermi 
distribution function with the corresponding muon chemical potential. 
These fully inelastic charged-current absorption rates are shown in
Fig.~\ref{fig:cc} (solid lines) for $\nu_\mu$ (left panel) and
$\bar\nu_\mu$ (right panel) at two selected conditions, in comparison
with the CC rates in the elastic approximation (see Appendix~\ref{sec:B1}).
For the elastic rates we include the approximate treatment of 
inelasticity and weak magnetism corrections~\cite{Horowitz:2001xf}
via ($\bar\nu_\mu$)$\nu_\mu$-energy dependent multiplicative factors.

\begin{table}[b!]
\centering
\caption{NMS vector and axial-vector coupling constants.}
\label{tab:cvca}
\begin{ruledtabular}
\begin{tabular}{crcr}
Scattering process & $C_V$\footnote{$\sin^{2}\theta_W\approx 0.23$} & & $C_A$ \\
\hline
$\nu_e + \mu^\pm$           &  $-0.5+2\sin^2\theta_W$ & & $\pm 0.5$ \\
$\bar\nu_e + \mu^\pm$     &  $-0.5+2\sin^2\theta_W$ & & $\mp 0.5$ \\
$\nu_\mu + \mu^\pm$       &  $ 0.5+2\sin^2\theta_W$ & & $\mp0.5$ \\
$\bar\nu_\mu + \mu^\pm$ &  $ 0.5+2\sin^2\theta_W$ & & $\pm0.5$ \\
$\nu_\tau + \mu^\pm$       &  $-0.5+2\sin^2\theta_W$ & & $\pm0.5$ \\
$\bar\nu_\tau + \mu^\pm$ &  $-0.5+2\sin^2\theta_W$ & & $\mp0.5$ \\
\end{tabular}
\end{ruledtabular}
\end{table}

For $\nu_\mu$ and $\bar\nu_\mu$ energies below the $Q$ values of 
$m_\mu-(m_n-m_p)-(U_n-U_p)$ and $m_\mu+(m_n-m_p)+(U_n-U_p)$, respectively,
there can be no contribution to the opacity within the elastic treatment
(details about the elastic rate expression are given in
Appendix~\ref{sec:B1}), as illustrated in Fig.~\ref{fig:cc} 
(dashed lines) at two selected conditions (a) and (b), which are 
listed in Table~\ref{tab:eos}. This strong opacity drop is modified 
when taking into account inelastic contributions within the full 
kinematics approach, as illustrated in  Fig.~\ref{fig:cc} (solid 
lines). Since these muonic CC processes are expected to be 
responsible for the muonization of SN matter, from 
Fig.~\ref{fig:cc} it becomes evident this channel requires 
$\nu_\mu$ of high energy in order to produce final state muons 
in non-negligible amounts.

\begin{figure*}[htp]
\subfigure[]{\includegraphics[width=1.0\columnwidth]{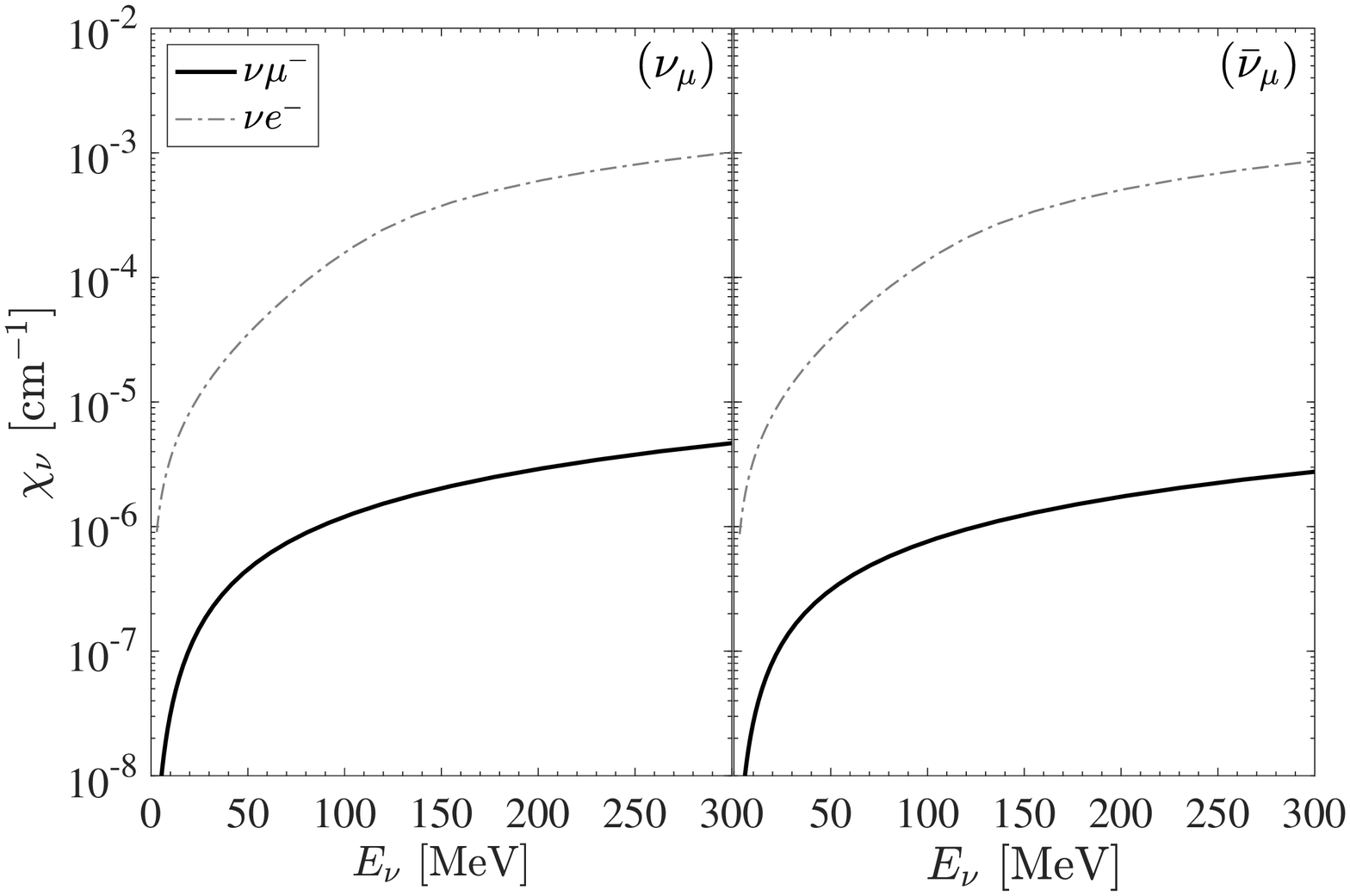}}
\hfill
\subfigure[]{\includegraphics[width=1.0\columnwidth]{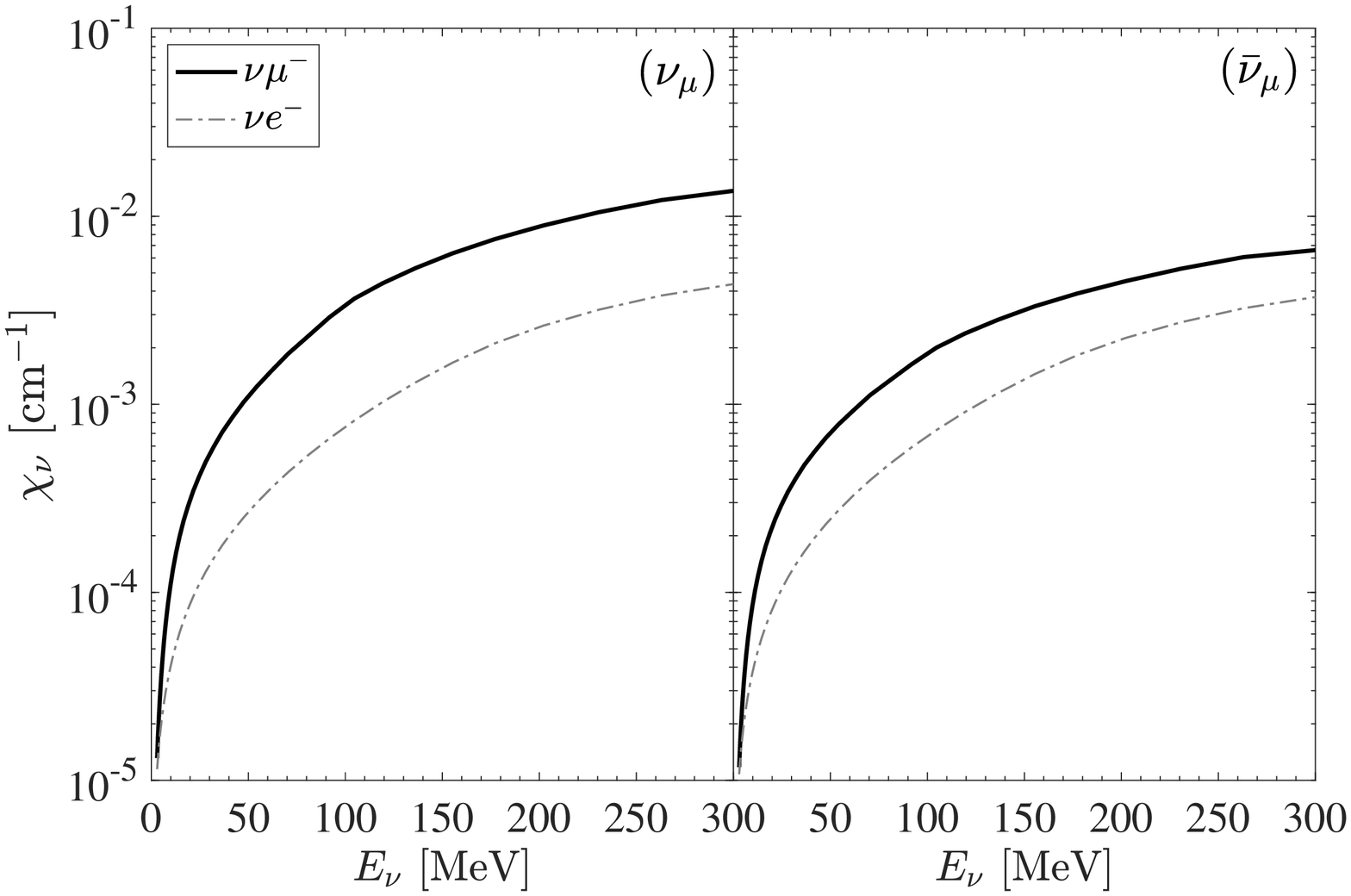}}
\caption{Neutrino scattering opacity, Eq.~\eqref{eq:opacity_reduced}, 
for $\nu_\mu$ (left panels) and $\bar\nu_\mu$ (right panels) on muons 
(solid lines) and on electrons (dash-dotted lines), in both cases 
assuming a free final-state neutrino according to 
expression~\eqref{eq:opacity_reduced}, at two selected conditions 
referred to as (a) and (b), which are listed in Table~\ref{tab:eos}.}
\label{fig:nes-nms}
\end{figure*}

\subsubsection{Neutrino--muon scattering}
For neutrino--muon scattering (NMS), reactions~(2a) and (2b) in 
Table~\ref{tab:muonic-weak}, the approach for neutrino--electron 
scattering (NES) is employed here following the detailed derivation 
provided in 
Refs.~\cite{Tubbs:1975jx,SchinderShapiro:1982,Bruenn:1985en,Mezzacappa:1993gx},
which is equivalent to the recent derivation in Ref.~\cite{Guo:2020},
Section~II~A and Appendix~A. Mapping the algebraic expressions of
NES to NMS is straightforward due to the similarity of the
transition amplitudes and hence of the scattering kernels between
NES and NMS. It requires the replacement of electron restmass
and chemical potential with those of the muons. However, the
vector and axial-vector coupling constants are different for
NES and NSM, which are listed in Table~\ref{tab:cvca}. Details
about the scattering amplitudes and in/out scattering kernels
for NMS, $\mathcal{R}_{\rm NMS}^{\rm in/out}$, are provided in
Appendix~\ref{sec:B2}, together with their implementation in
the collision integral of {\tt AGILE-BOLTZTRAN}.

\begin{figure*}[ht!]
  \subfigure[]{\includegraphics[width=0.975\columnwidth]{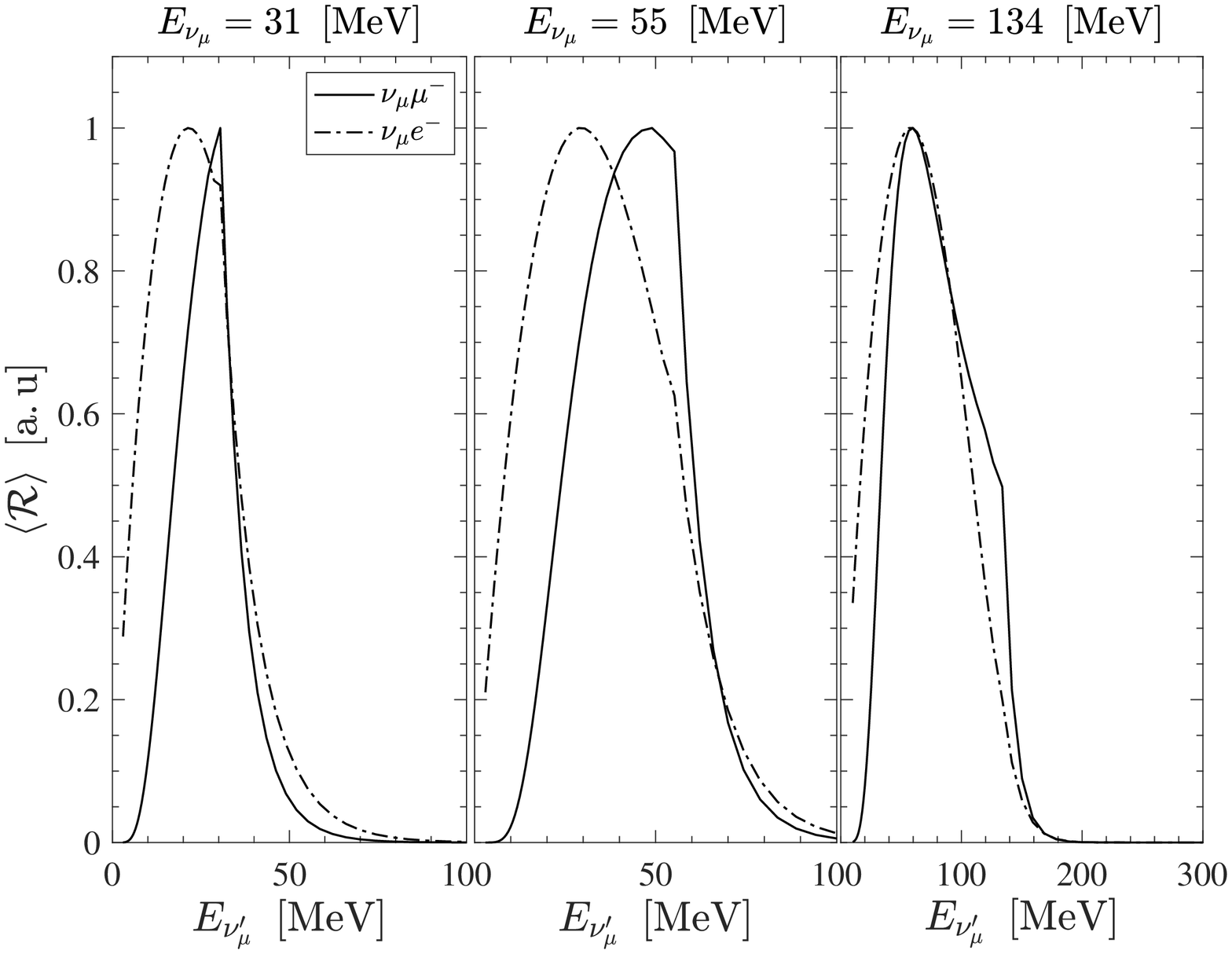}
  \label{fig:R_individual_nes-nms_a}}
\hfill
\subfigure[]{\includegraphics[width=0.975\columnwidth]{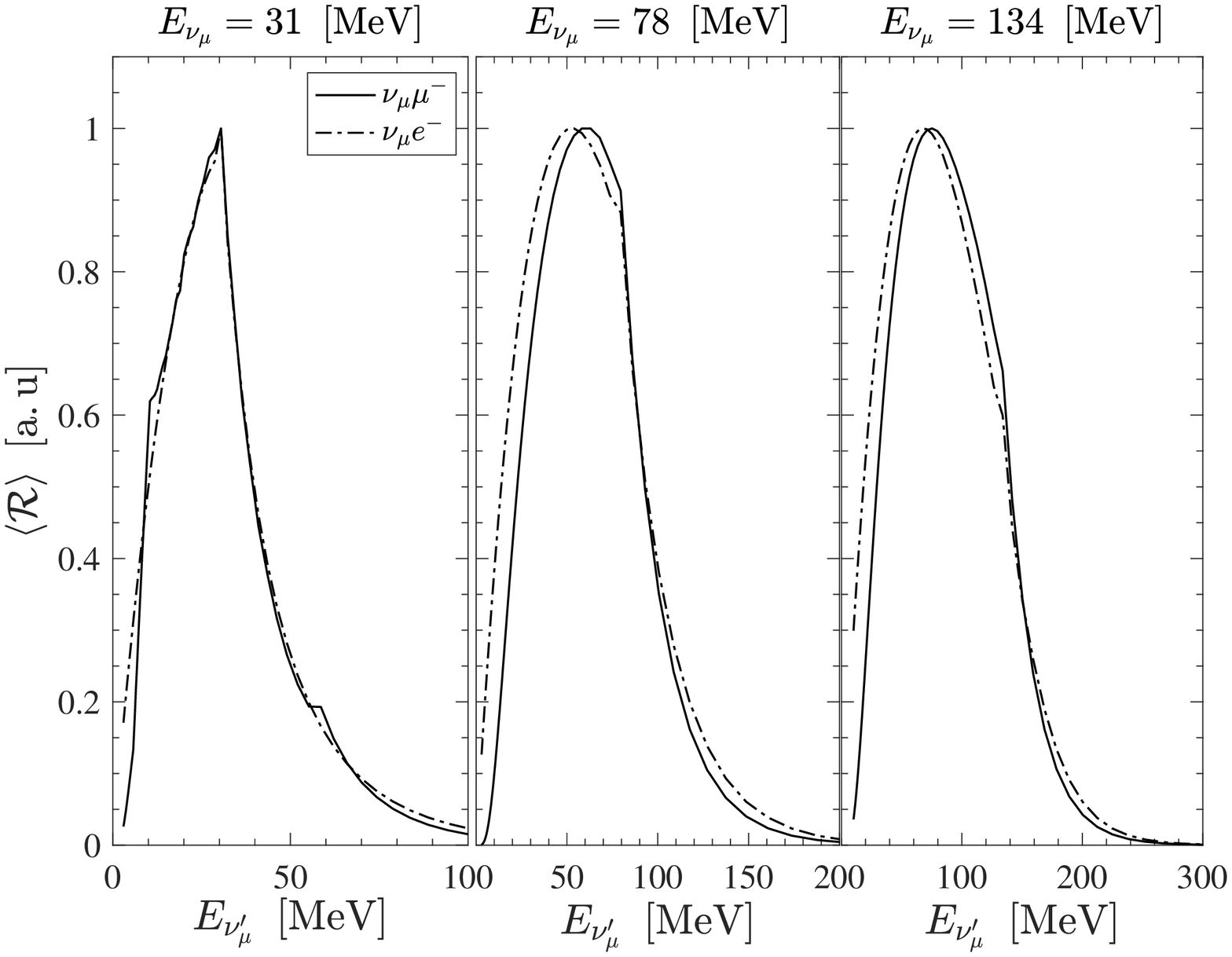}
  \label{fig:R_individual_nes-nms_b}}
\caption{Angular averaged out-scattering kernel, $\langle \mathcal{R} \rangle$
Eq.~\eqref{eq:R_intdmu}, normalized to unity, for three different incoming 
neutrino energies, $E_\nu$, comparing NMS (solid lines) and NES 
(dash-dotted lines) as a function of the out-going neutrino energy, 
$E_{\nu'}$, for the conditions (a) and the conditions (b) according to 
Table~\ref{tab:eos}.}
\label{fig:R_individual_nes-nms}
\end{figure*}

In Fig.~\ref{fig:nes-nms} we compare the opacity for neutrino--$e^\pm$
scattering (dashed lines) and neutrino--$\mu^\pm$ scattering (solid lines) 
at two selected conditions (a) and (b), for which the thermodynamic 
conditions are listed in Table~\ref{tab:eos}. In order to obtain the 
neutrino-scattering opacity, the following integration is performed 
of the out-scattering kernel, $\mathcal{R}_{\rm NMS}^{\rm out}$, over 
the final-state neutrino phase space,
\begin{eqnarray}
\chi_{\nu}(E_{\nu}) &=& \frac{1}{(2\pi\hbar c)^3}\int dE_{\nu'}E_{\nu'}^2  \int d\left(\cos\vartheta'\right)  \int d\left(\cos\vartheta\right) \nonumber \\
&& \; \times \;  \mathcal{R}^{\rm out}(E_{\nu},E_{\nu'},\vartheta,\vartheta')~;
\label{eq:opacity_reduced}
\end{eqnarray}
i.e. assuming a free-final state neutrino phase space (more details can
be found in Ref.~\cite{Fischer:2012a}). The scattering kernel depends
on the in-coming and the out-going neutrino energies, $E_\nu$ and
$E_\nu'$, as well as the in-coming and out-going relative angles,
$\vartheta$ and $\vartheta'$ (see Appendix~\ref{sec:B}).

Note that neutrino trapping and thermalization of $\mu$ and $\tau$
neutrinos occurs roughly at the conditions between (a) and (b) of
Table~\ref{tab:eos}. Hence, neutrino--muon scattering may be an
important source for the thermalization and trapping of heavy
lepton-flavor neutrinos. Furthermore, from the comparison in
Fig.~\ref{fig:nes-nms} it becomes evident that at high densities,
muon-neutrino scattering on muons dominates over scattering on
electrons. This is mostly attributed to the high electron
degeneracy due to which the final-state electron phase space is 
occupied. Note also that, at such conditions (b), neutrinos are 
trapped.

\begin{figure*}[ht!]
  \subfigure[~$E_{\nu_\mu}=31$~MeV, NMS:~$E_{\nu_\mu'}=31$~MeV, NES:~$E_{\nu_\mu'}=24$~MeV]
  {\includegraphics[width=0.98\columnwidth]{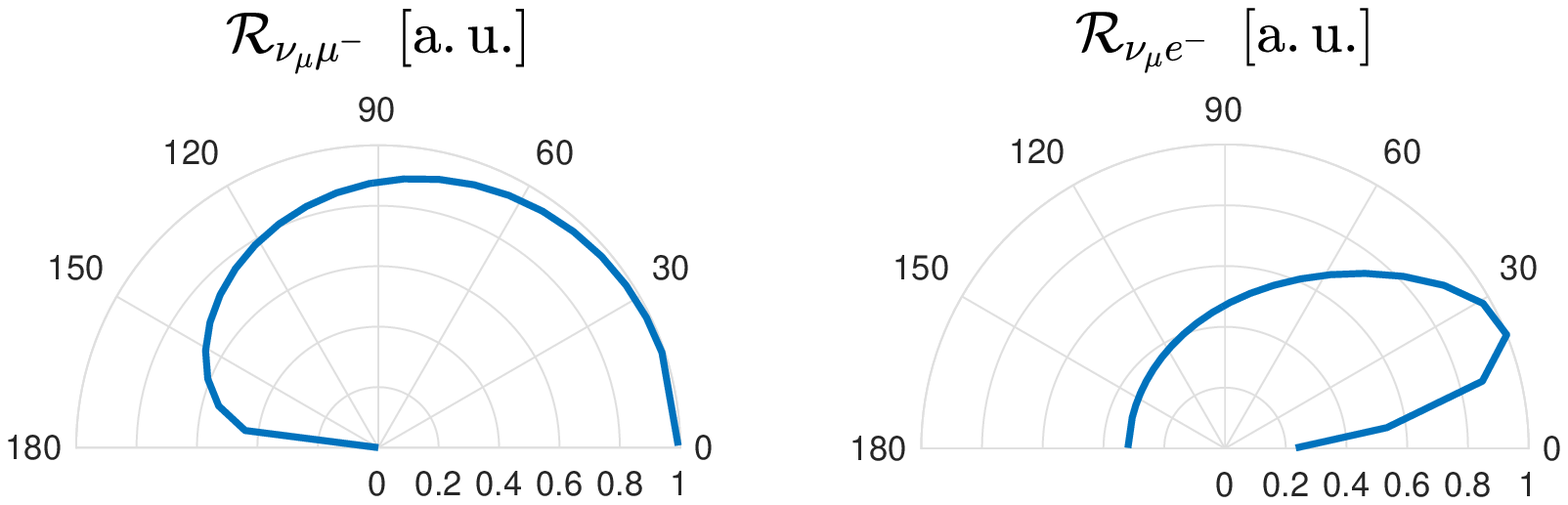}
   \label{fig:R_individual_angular_nes-nms_a}}
   \hspace{5mm}
  \subfigure[~$E_{\nu_\mu}=78$~MeV, NMS:~$E_{\nu_\mu'}=60$~MeV, NES:~$E_{\nu_\mu'}=51$~MeV]
  {\includegraphics[width=0.98\columnwidth]{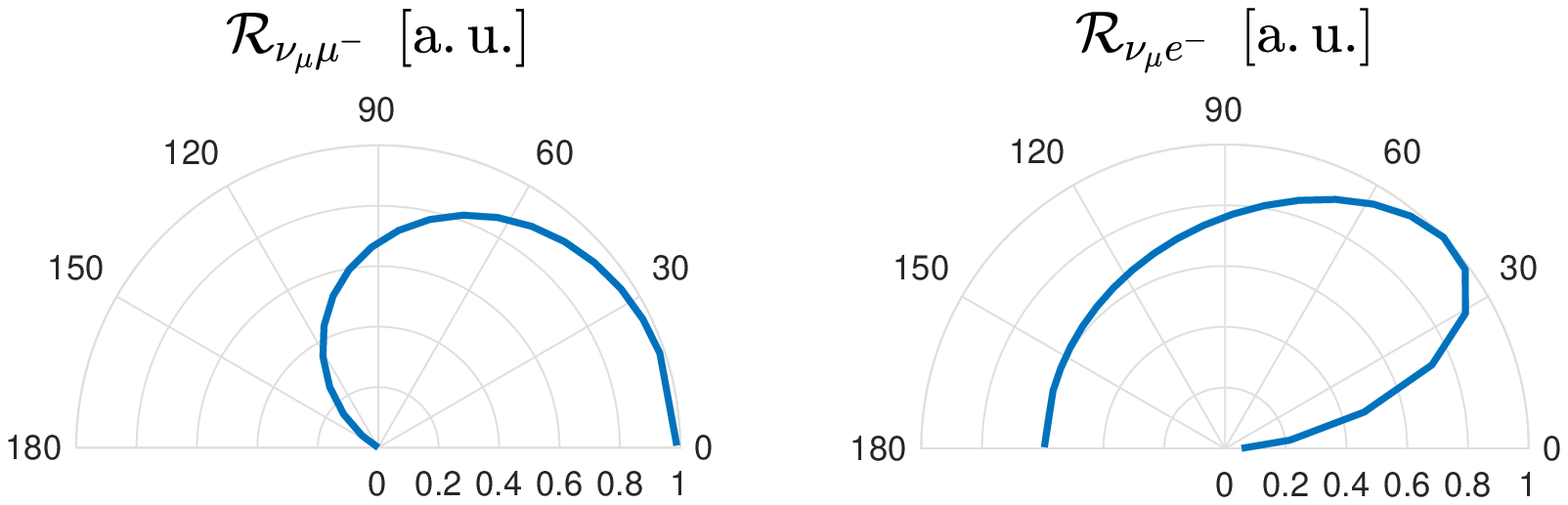}
   \label{fig:R_individual_angular_nes-nms_b}}
\\
\vspace{7mm}
  \subfigure[~$E_{\nu_\mu}=31$~MeV, NMS:~$E_{\nu_\mu'}=35$~MeV, NES:~$E_{\nu_\mu'}=35$~MeV]
  {\includegraphics[width=0.98\columnwidth]{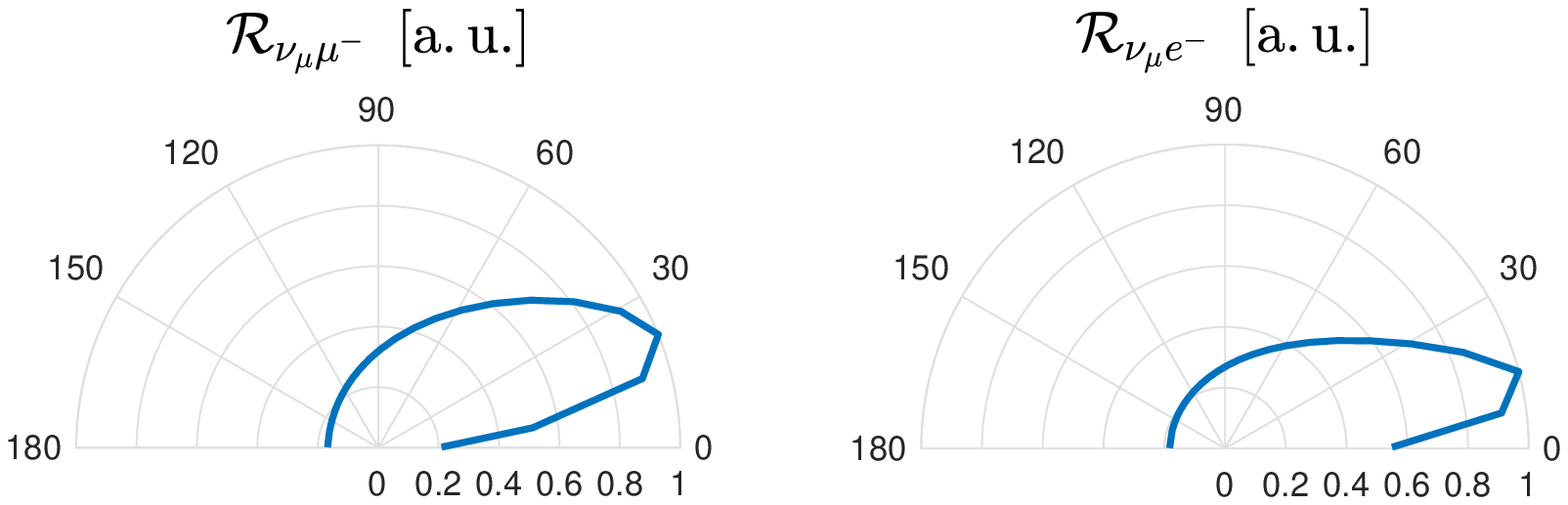}
  \label{fig:R_individual_angular_nes-nms_c}}
  \hspace{5mm}
  \subfigure[~$E_{\nu_\mu}=78$~MeV, NMS:~$E_{\nu_\mu'}=100$~MeV, NES:~$E_{\nu_\mu'}=100$~MeV]
  {\includegraphics[width=0.98\columnwidth]{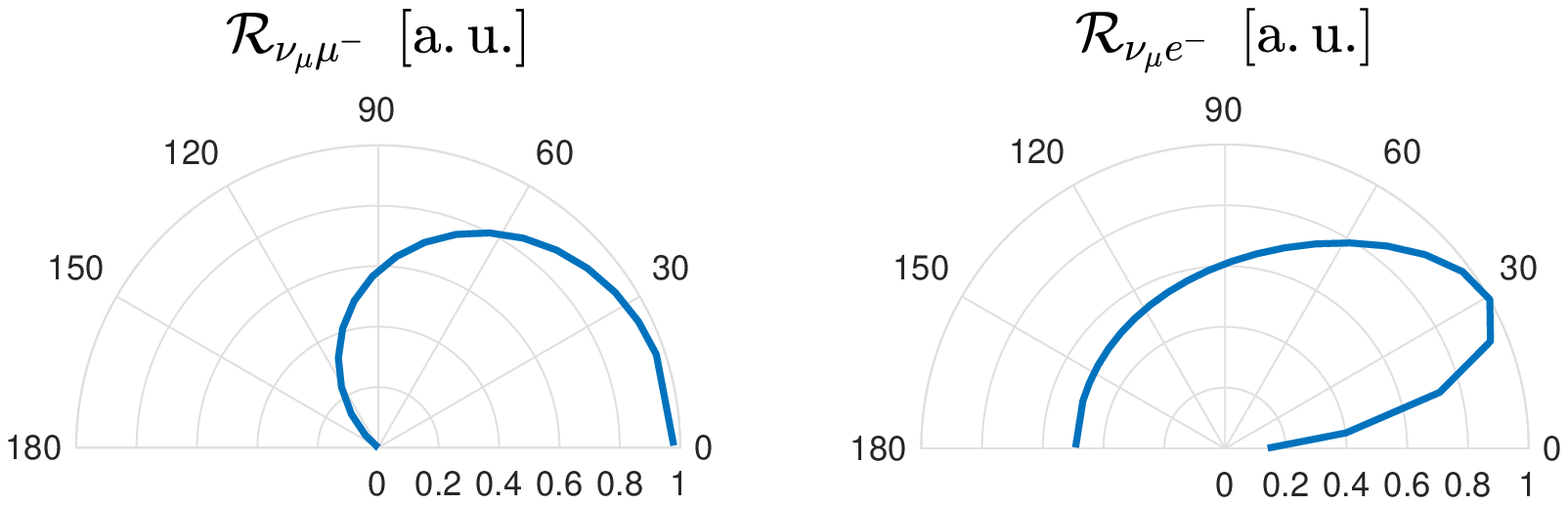}
  \label{fig:R_individual_angular_nes-nms_d}}
\\
\vspace{7mm}
  \subfigure[~$E_{\nu_\mu}=31$~MeV, NMS:~$E_{\nu_\mu'}=18$~MeV, NES:~$E_{\nu_\mu'}=8$~MeV]
  {\includegraphics[width=0.98\columnwidth]{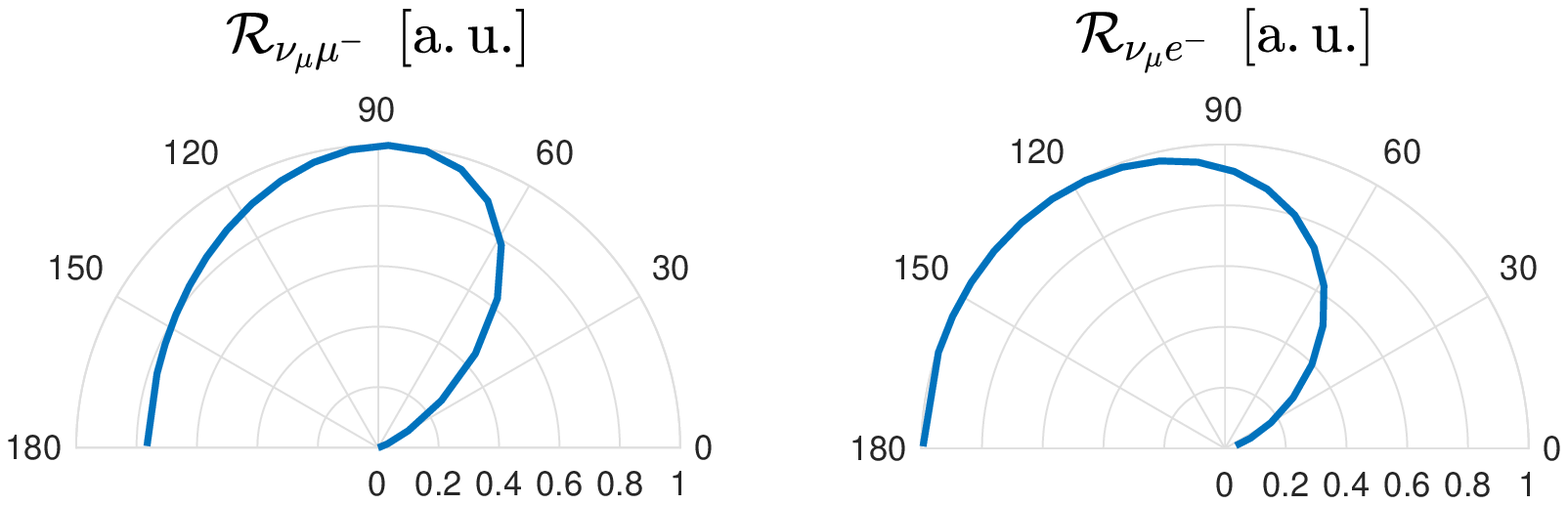}
  \label{fig:R_individual_angular_nes-nms_e}}
  \hspace{5mm}
  \subfigure[~$E_{\nu_\mu}=78$~MeV, NMS:~$E_{\nu_\mu'}=22$~MeV, NES:~$E_{\nu_\mu'}=8$~MeV]
  {\includegraphics[width=0.98\columnwidth]{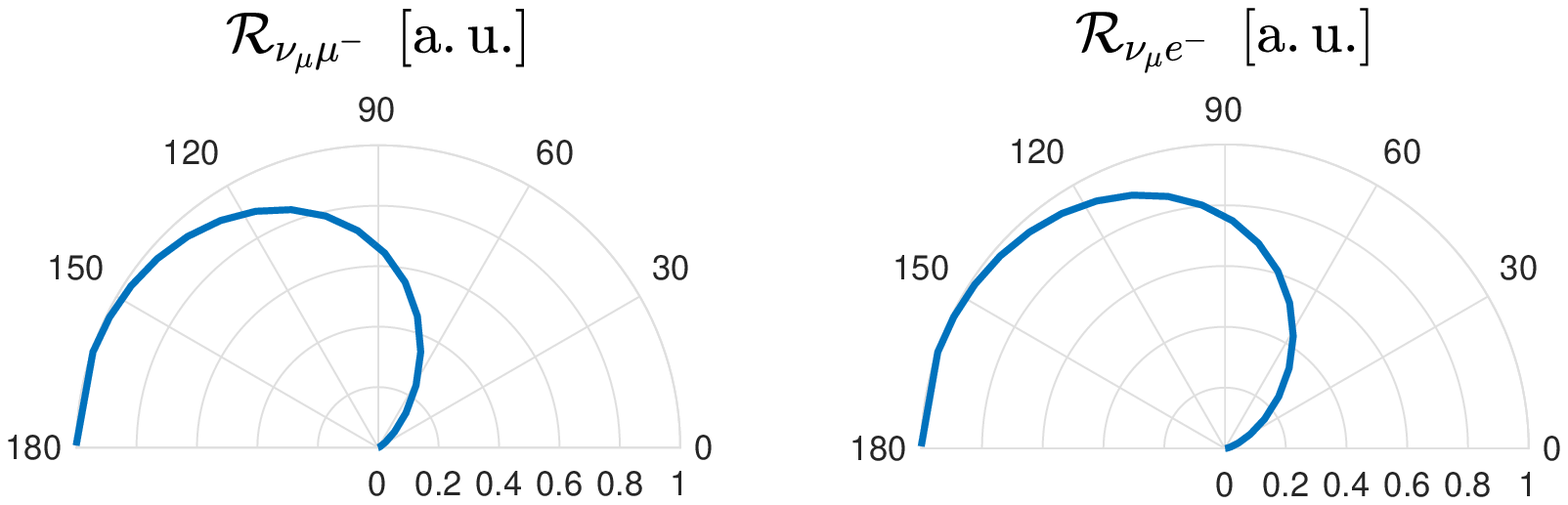}
  \label{fig:R_individual_angular_nes-nms_f}}
\caption{Angular distribution for $\theta$ in degrees ($\cos\theta$ is defined in 
Eq.~\eqref{eq:costheta}) of the out-scattering kernel, $\mathcal{R}(E_\nu,E_\nu',\cos\theta)$,
in arbitrary units, comparing $\nu_\mu$ scattering on muons, $\mathcal{R}_{\nu_\mu \mu^-}$,
and on electrons, $\mathcal{R}_{\nu_\mu e^-}$, for incoming neutrino energies,
$E_{\nu_\mu}= 3.15\,T$; i.e. $E_{\nu_\mu}=31$~MeV for $T=10$~MeV (left panels) 
and $E_{\nu_\mu}=78$~MeV for $T=25$~MeV (right panels), for the conditions~(a)
of in Table~\ref{tab:eos} in \ref{fig:R_individual_angular_nes-nms_a}, 
\ref{fig:R_individual_angular_nes-nms_c} and 
\ref{fig:R_individual_angular_nes-nms_e}, and the conditions~(b) of 
Table~\ref{tab:eos} in \ref{fig:R_individual_angular_nes-nms_b}, 
\ref{fig:R_individual_angular_nes-nms_d} and 
\ref{fig:R_individual_angular_nes-nms_f}, and varying out-scattering energies
$E_{\nu_\mu'}$ corresponding to the peak of the scattering kernel and the half-width
(see text for details).}
\label{fig:R_individual_angular_nes-nms}
\end{figure*}

Since the opacity shown in Fig.~\ref{fig:nes-nms} does not reveal insights 
into the inelasticity of the processes, in Fig.~\ref{fig:R_individual_nes-nms}
we show in addition the angular-averaged outgoing scattering kernels defined
as follows,
\begin{eqnarray}
\langle \mathcal{R} \rangle(E_{\nu},E_{\nu'}) &=& \frac{1}{(2\pi\hbar c)^3} E_{\nu_\mu'}^2 \int d\left(\cos\vartheta'\right)  \int d\left(\cos\vartheta\right) \nonumber \\
&& \; \times \; \mathcal{R}^{\rm out}(E_{\nu},E_{\nu'},\vartheta,\vartheta')~,
\label{eq:R_intdmu}
\end{eqnarray}
as a function of the outgoing neutrino energy, $E_{\nu'}$, for three different 
incoming neutrino energies, $E_\nu$, evaluated at the two conditions (a) and (b)
listed in Table~\ref{tab:eos} shown in Figs.~\ref{fig:R_individual_nes-nms_a}
and \ref{fig:R_individual_nes-nms_b}, respectively. 
Low incoming neutrino energies (left panels in Fig.~\ref{fig:R_individual_nes-nms})
both NMs (solid lines) and NES (dashed-dotted lines) are dominated by 
down-scattering, due to the electrons being degenerate and, since muons are
never degenerate under SN conditions, the high muon restmass. With increasing
$E_{\nu_\mu}$ (middle and right panels in Fig.~\ref{fig:R_individual_nes-nms})
the restmass contribution becomes less important, and the differences between
NMS and NES are due to the different degeneracy.

In addition, Figure~\ref{fig:R_individual_angular_nes-nms} 
shows the dependence of the scattering kernel on the total scattering
angle, $\cos\theta$ for the conditions~(a) in Table~\ref{tab:eos} (left panels)
and the conditions~(b) in Table~\ref{tab:eos} (right panels),
assuming thermal energies for the initial neutrinos; i.e.,
$E_\nu=E_{\nu'}=3.15\,T$ and for different out-going neutrino energies
$E_{\nu_\mu'}$
Therefor, the top panel in Fig.~\ref{fig:R_individual_angular_nes-nms}
assumes out-scattering energies which are equal to the peak of the
scattering kernel; i.e. $E_{\nu_\mu'}=31$~MeV (NMS) and $E_{\nu_\mu'}=24$~MeV
(NES) for $E_{\nu_\mu}=31$~MeV corresponding to conditions~(a) in Table~\ref{tab:eos}
($T=10$~MeV), from where it becomes evident that NES
is mainly forward peaked at an angle of about 30~degrees while NMS is more
isotropic. Note that the scale in Figs.~ref\ref{fig:R_individual_angular_nes-nms_a}
and \ref{fig:R_individual_angular_nes-nms_b} are logarithmic.
At the conditions illustrated in Fig.~\ref{fig:R_individual_angular_nes-nms_c},
which correspond to out-scattering energies being equal to the peak of the
scattering kernels plus the half-width (see the left panel in
Fig.~\ref{fig:R_individual_nes-nms_a}), both NMS and NES are strongly
forward peaked. This situation is reversed in Fig.~\ref{fig:R_individual_angular_nes-nms_e}
where the out-scattering energy is equal to the peak of the scattering kernel
minus the half-width, when both NMS and NES are back-scattering dominated
with NMS being more isotropic.
This situation remains the same at higher density and $E_{\nu_\mu}=78$~MeV ($T=25$~MeV) 
orresponding to the conditions (b) in Table~\ref{tab:eos}, illustrated in
Figs.~\ref{fig:R_individual_angular_nes-nms_b}
\ref{fig:R_individual_angular_nes-nms_d} and \ref{fig:R_individual_angular_nes-nms_f}
(for the scattering kernel, see the middle panel in Fig.~\ref{fig:R_individual_nes-nms_b}).

\subsubsection{Purely leptonic reactions -- (i) lepton flavor exchange}
A new class of weak processes, known as lepton flavor exchange 
(LFE) reactions~\cite{Bollig:2017}, is added to {\tt AGILE-BOLTZTRAN}, 
reactions (3a) and (3b) of Table~\ref{tab:muonic-weak}. The 
close analogy of the scattering amplitudes of NMS and LFE, enables 
the direct comparison between these two processes, which simplifies 
the calculation of the in- and out-scattering kernels, provided in 
detail in Appendix~\ref{sec:B3a} where the nomenclature of 
Ref.~\cite{Mezzacappa:1993gx} is followed closely. It is 
equivalent to the recent derivation in Ref.~\cite{Guo:2020}, 
Section~II~A and Appendix~A.

The main difference between NMS and LFE is the appearance of a new
energy scale since initial- and final-state leptons are different; 
one has to take the restmass energy difference between muon and 
electron into account. This gives rise to additional terms in the 
scattering amplitudes (they are provided in Appendix~\ref{sec:B3a}), 
which can be large.

Figure~\ref{fig:leptonic} compares the set of the LFE processes~(3a)
and (3b) in Table~\ref{tab:muonic-weak}, for each channel
individually at the two selected conditions (a) and (b) 
corresponding to Table~\ref{tab:eos}. For the calculation of the
opacity the same approach is implemented here as for neutrino--muon
scattering Eq.~\eqref{eq:opacity_reduced}; i.e., assuming a free
final-state neutrino. These rates are in agreement with those
obtained in Ref.~\cite{Guo:2020} with a detailed comparison of the
LFE rates and the muonic CC rates. 

\begin{figure*}[t!]
\centering
\subfigure[]
{\includegraphics[width=1.0\columnwidth]{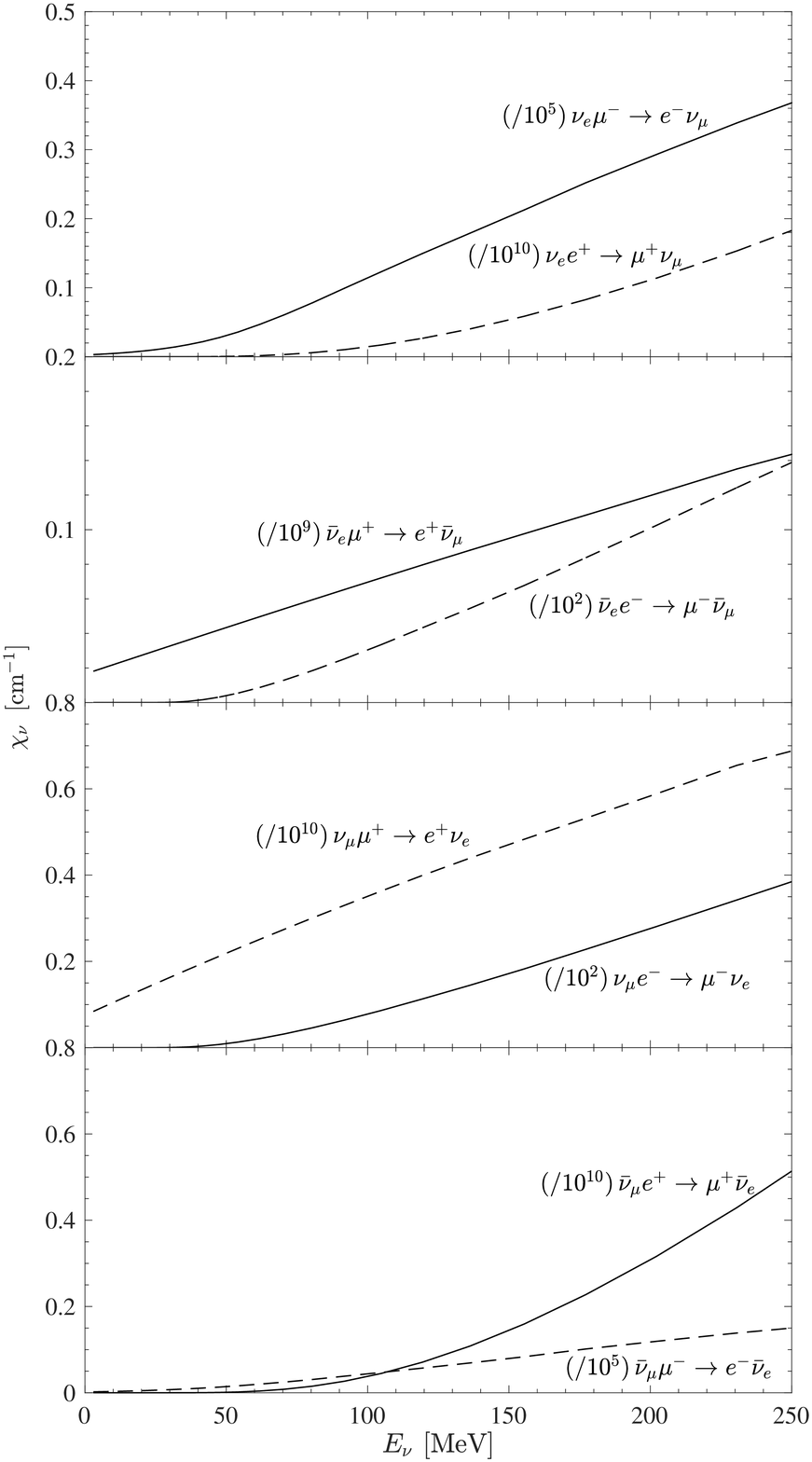}\label{fig:leptonic_a}}
\hfill
\subfigure[]
{\includegraphics[width=1.0\columnwidth]{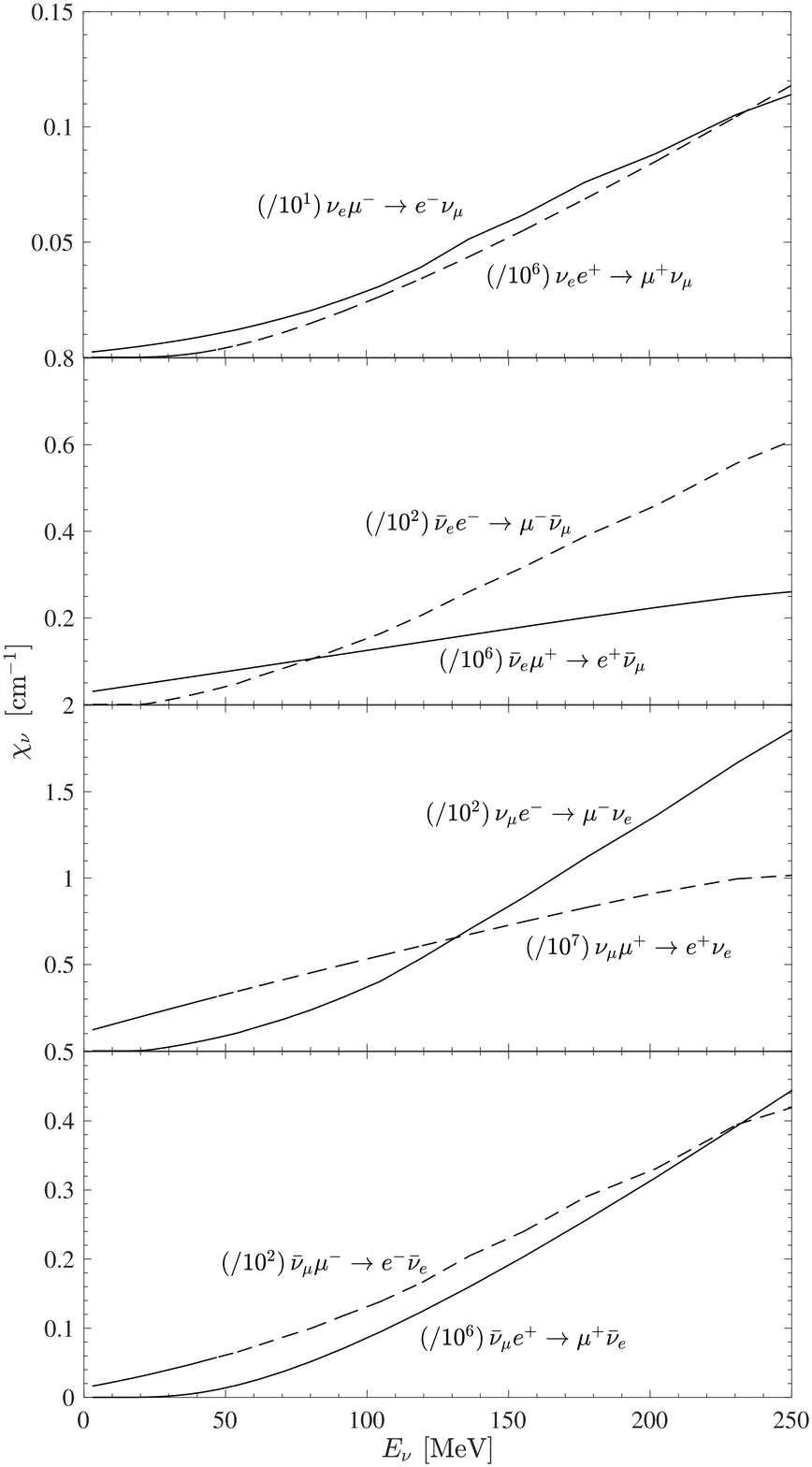}\label{fig:leptonic_b}
}
\caption{Opacity for the purely leptonic processes, lepton-flavor
exchange~(solid lines) and lepton-flavor conversion~(dashed lines), 
assuming a free final state neutrino, at the same two selected conditions 
labelled (a) and (b) for which the thermodynamic state is listed in 
Table~\ref{tab:eos}.}
\label{fig:leptonic}
\end{figure*}

\subsubsection{Purely leptonic reactions -- (ii) lepton flavor conversion}
There is a second class of purely leptonic processes involving 
(anti)muons, known as lepton flavor conversion reactions 
(LFC)~\cite{Bollig:2017}, reactions (4a) and (4b) in 
Table~\ref{tab:muonic-weak}. The derivation of the in- and 
out-scattering kernels is given in Appendix~\ref{sec:B3b}, 
again in close analogy to Ref.~\cite{Guo:2020}. 
Figure~\ref{fig:leptonic} compares the rates for the LFC 
processes, at the same two selected conditions (a) and (b) 
of Table~\ref{tab:eos}, as before. This comparison is in 
agreement with the analysis of Ref.~\cite{Guo:2020}. 

All these weak reaction rates involving muons, i.e. muonic CC
rates and NMS as well as the LFE and LFC processes, are
implemented in {\tt AGILE-BOLTZTRAN} within the 6-species setup,
in order to simulate and study the impact of the muonization of
SN matter. In the following, these results will be discussed as
the reference case and compared to the simulations where all muonic
weak rates are set to zero. Note that we omit here the (inverse)
muon decay.

\section{Core-collapse SN simulations}
\label{sec:sim}
The core-collapse SN simulations discussed in the following are
launched from the 18~M$_\odot$ progenitor from the stellar evolution
series of Ref.~\cite{Woosley:2002zz}. Besides the muonic weak
processes introduced in Sec.~\ref{sec:SNmodel} above, the standard
set of non-muonic weak reactions employed here is given in
Table~(1) of Ref.~\cite{Fischer:2020}. A comparison of these
non-muonic weak rates in the `minimal' setup of
S.~Bruenn~\cite{Bruenn:1985en} and major 
updates~\cite{Horowitz:2001xf,Juodagalvis:2010,MartinezPinedo:2012,Roberts:2012a,Fischer:2016a},
including the impact in spherically-symmetric and
axially-symmetric SN simulations, is provided in
Refs.~\cite{Lentz:2012,Kotake:2018a}.

\subsection{Production of muons at core bounce}
\label{sec:muoniztion}

\begin{figure}[t!]
\includegraphics[width=0.9975\columnwidth]{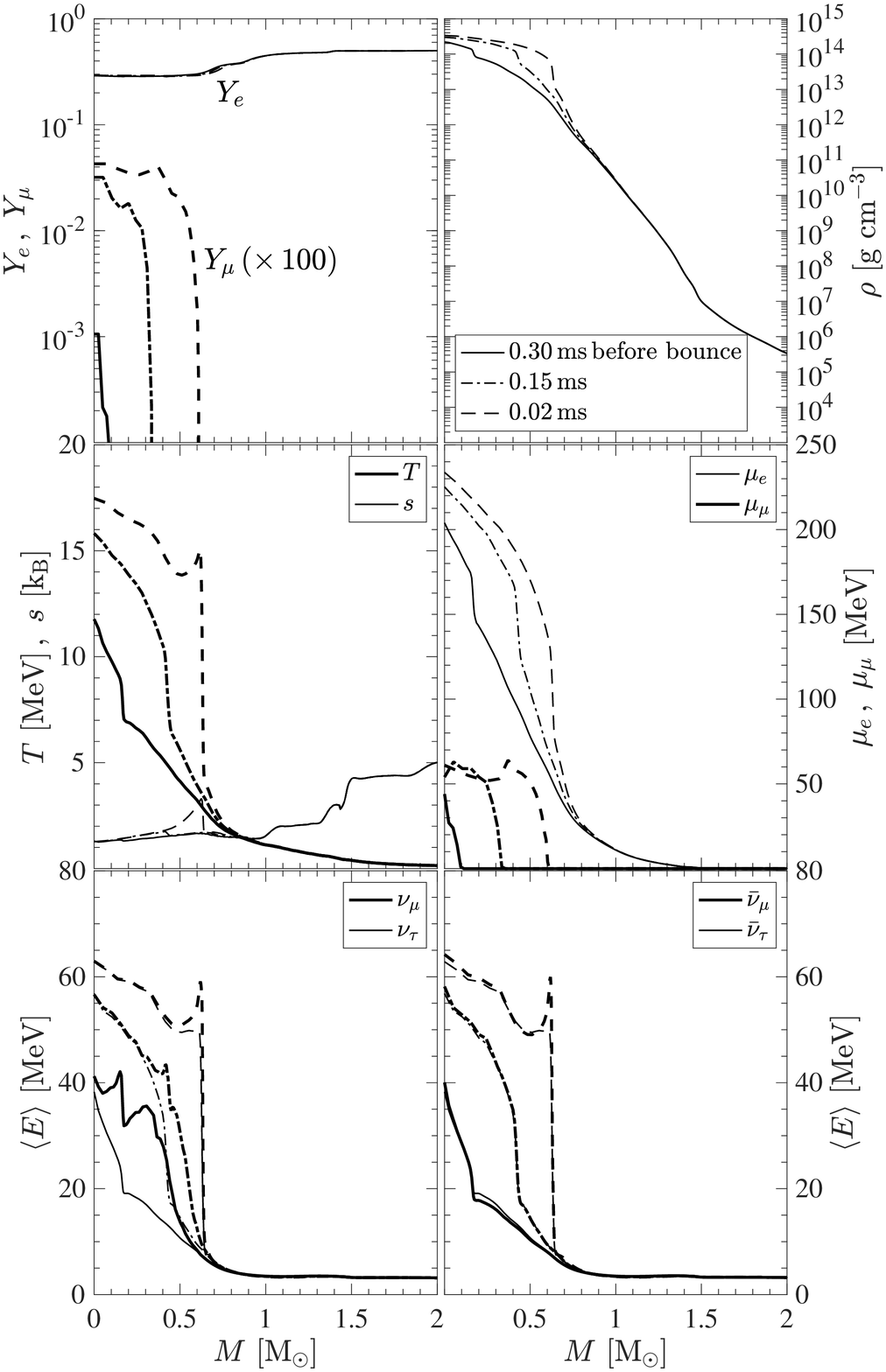}
\caption{(Color online) Radial profiles of selected quantities
as a function of the enclosed baryon mass, showing the electron
and muon fractions ($Y_e$, $Y_\mu$), restmass density ($\rho$),
temperature and entropy per baryon ($T$, $s$), electron and
muon chemical potentials ($\mu_e$,$\mu_\mu$), as well as the
the average neutrino energies for $\mu$ and $\tau$
(anti)neutrino flavors. The conditions correspond to a few
tenths of a millisecond before core bounce. The CC rates employed here are
within the full kinematics treatment}
\label{fig:hydro-collapse}
\end{figure}

\begin{figure}[t!]
\includegraphics[width=0.9975\columnwidth]{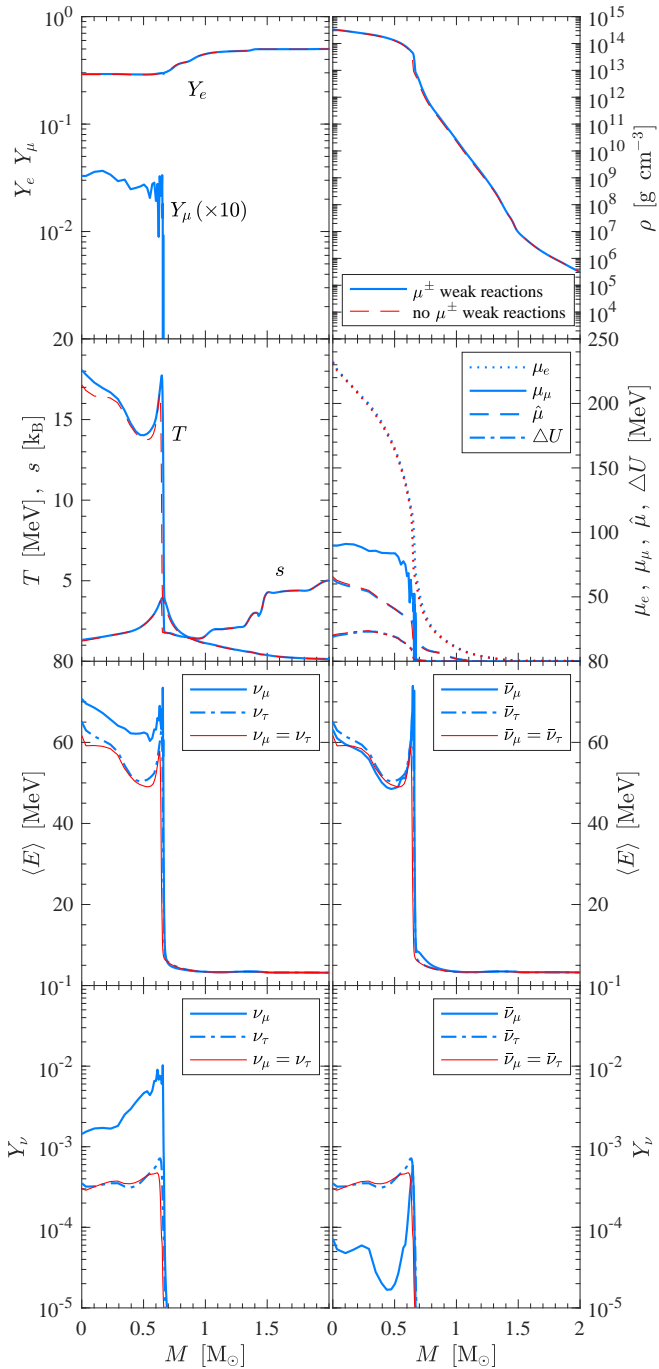}
\caption{(Color online) The same quantities are shown as in Fig.~\ref{fig:hydro-collapse} but at core bounce, comparing
the reference simulation with (blue lines) and without muonic
weak processes (red lines). In addition to the electron and muon 
chemical potentials, we show the charged chemical potential denoted
as $\hat\mu=\mu_n-\mu_p$ and the mean-field potential difference
$\triangle U=U_n-U_p$, as well as the neutrino abundances $Y_\nu$
for $\mu-$ and $\tau-$(anti)neutrinos.}
\label{fig:hydro-bounce}
\end{figure}

There could be two mechanisms for the production of muons. One is 
driven by electromagnetic pair processes, such as
$e^- + e^+ \rightarrow \mu^- + \mu^+$, which are fast but require
high temperatures that are not reached in the simulation.
Furthermore, this process would always result in a zero net muon abundance.
The second mechanism is due to weak processes starting from the
production of muon (anti)neutrinos that are converted later
into muons. The latter is the dominant channel here. Furthermore,
due to the largely different CC opacity for $\nu_\mu$ and
$\bar\nu_\mu$ a net muonic abundance can be created. However,
the muonic CC processes can only operate once a large enough
fraction of high-energy $\nu_\mu$ and $\bar\nu_\mu$ are produced,
which only occurs shortly before bounce. The origin of
high-energy muon-(anti)neutrinos are pair processes, mainly
electron--positron annihilation, when the temperature is
sufficiently high that positrons are present in the stellar
plasma, and $N$--$N$ bremsstrahlung processes.
This situation is illustrated in Fig.~\ref{fig:hydro-collapse}
(bottom panels) at a few tenths of a millisecond before core
bounce, when the average energies for $\nu_\mu$ and $\bar\nu_\mu$
reach as high as 50--70~MeV, due to temperatures on the order
of about 15~MeV. For the leading weak processes that give rise
to the muonization, the CC reactions~(1a) and (1b) in
Table~\ref{tab:muonic-weak}, the medium modifications
of the mean-field potentials can be as high as
$U_n-U_p\simeq20-30$~MeV (see Fig.~\ref{fig:hydro-bounce}).
This, in turn, enables the net-production of muons when the
average energy of the muon-neutrinos is substantially lower
than the muon restmass energy (see therefore
expression~\eqref{eq:EvEmu} in Appendix~\ref{sec:B1}
corresponding to the elastic rate approximation). Already at
core bounce this leads to a non-negligible muon abundance on
the order of $Y_\mu\simeq 10^{-3}$ (blue lines in
Fig.~\ref{fig:hydro-bounce}), in comparison to the simulation
setup without muonic weak processes (red lines). This, in turn,
feeds back to substantially different neutrino abundances $\nu_\mu$
and $\bar\nu_\mu$ which is not observed for the simulation without
muonic weak processes (see therefore the bottom panels in 
Fig.~\ref{fig:hydro-bounce}), where the origin of differences between
$\nu_\mu$ and $\bar\nu_\mu$ originates from different coupling
constants in neutrino-electron scattering. 
On the other hand here, the large difference between the abundances
of $\nu_\mu$ and $\bar\nu_\mu$ with $Y_{\nu_\mu}>Y_{\bar\nu_\mu}$
indicates the net muonization; i.e a substantially
higher abundance of $\mu^-$ then $\mu^+$.
Otherwise the evolution is in quantitative agreement with the simulation
without muonic weak processes, since such a muon abundance has a
negligible impact on the PNS structure (see Fig.~\ref{fig:hydro-bounce}).

\begin{figure*}[htp]
\subfigure[~At 1.7~ms (solid black lines) and 5~ms (grey dashed lines)]{
\includegraphics[width=1.0\columnwidth]{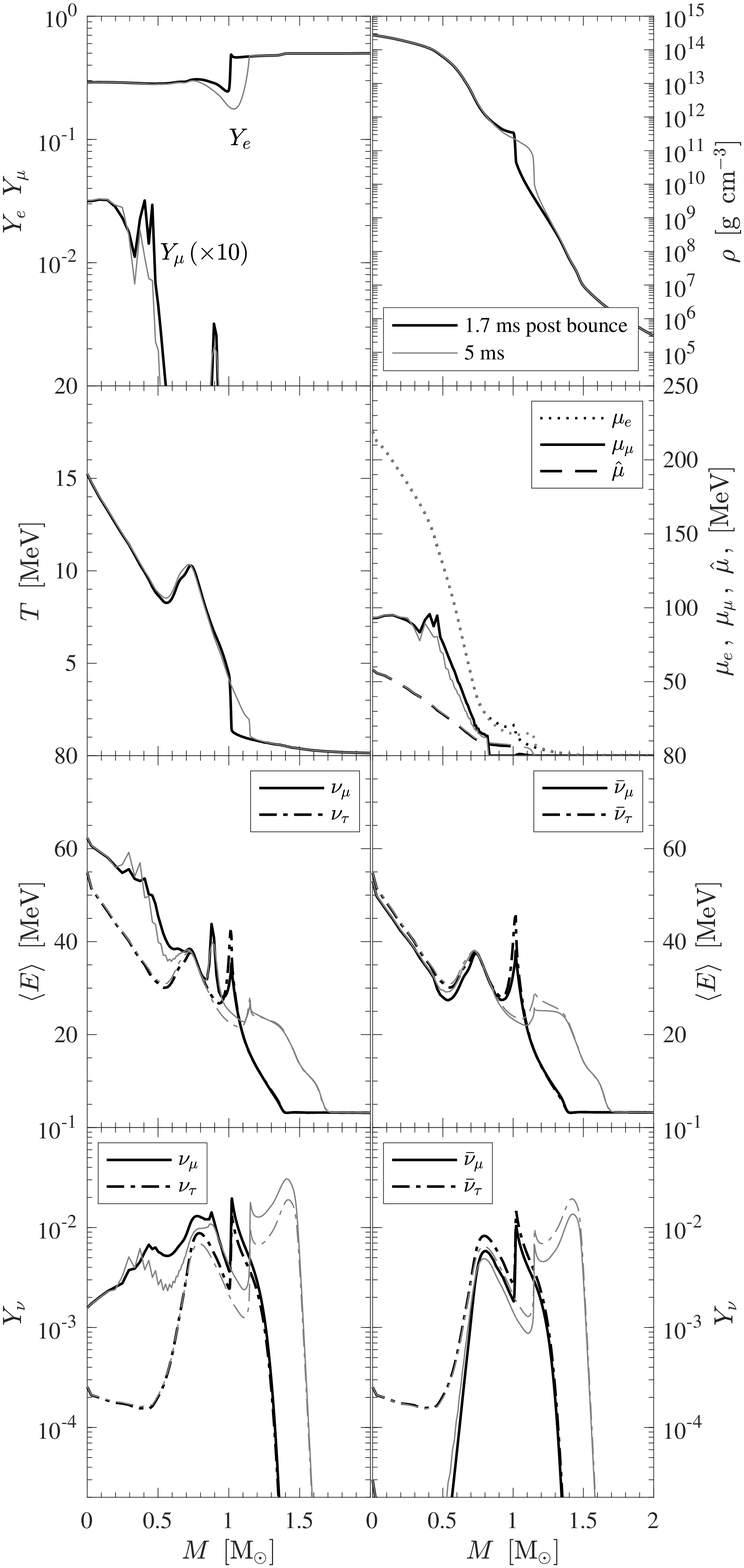}\label{fig:hydro-post-bounce_a}}
\hfill
\subfigure[~At 30~ms post bounce]{
\includegraphics[width=0.995\columnwidth]{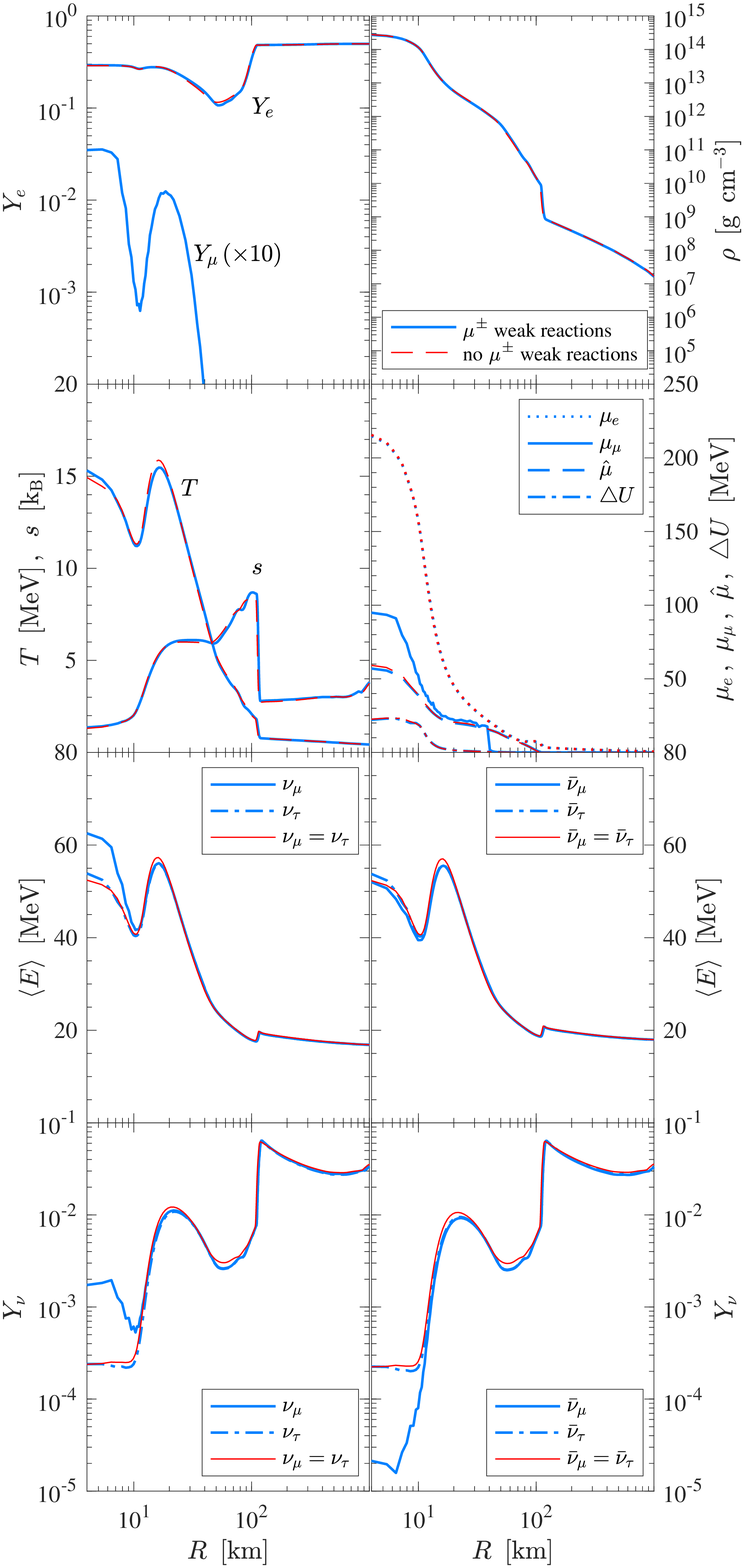}\label{fig:hydro-post-bounce_b}}
\caption{(Color online) The same quantities as in Fig.~\ref{fig:hydro-bounce}
for the simulation with muonic weak processes at selected times shortly
after the shock breakout in graph~(a) as a function of the enclosed
baryon mass, and at about 30~ms post bounce in graph~(b) as a
function of the radius. In addition in graph~(b) we compare the PNS structure
of the simulation with muonic weak reactions (blue lines) and
without muonic weak reaction (red lines).}
\label{fig:hydro-post-bounce}
\end{figure*}
\begin{figure*}[htp]
\subfigure[~Luminosty]{
\includegraphics[width=0.985\columnwidth]{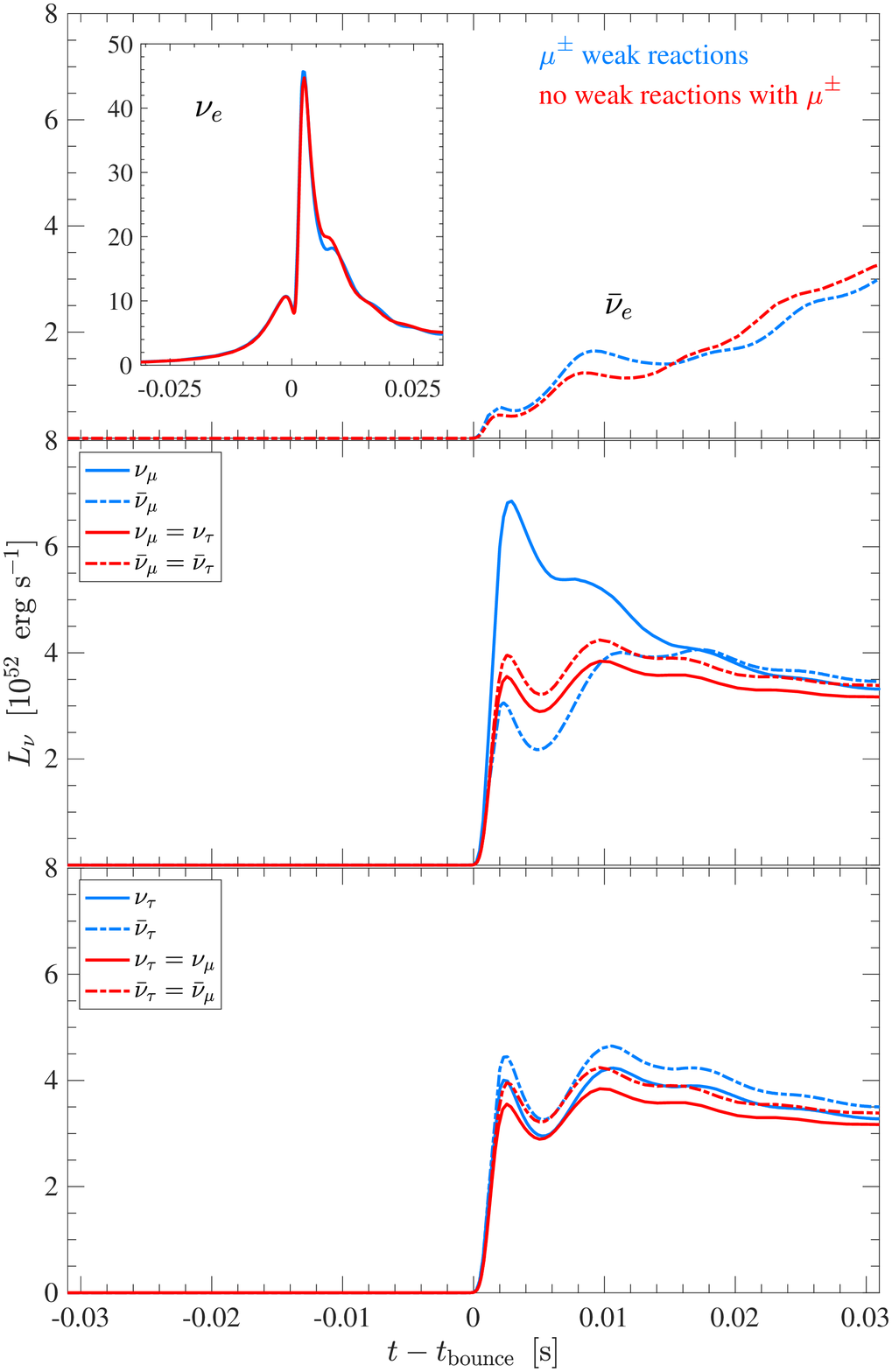}\label{fig:lumin_muonization}
}
\hfill
\subfigure[~Mean energy]{
\includegraphics[width=0.985\columnwidth]{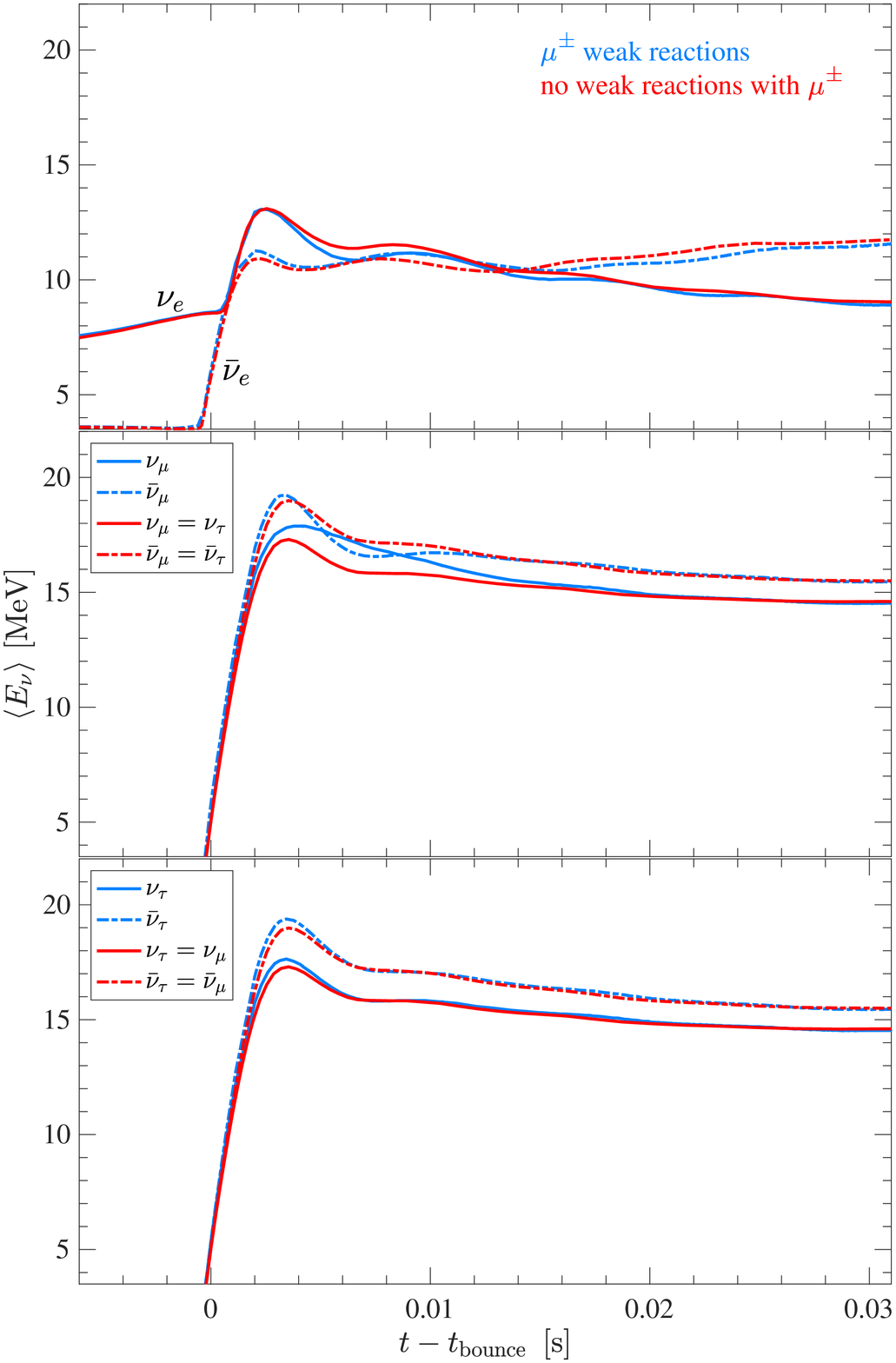}\label{fig:rms_muonization}

}
\caption{Evolution of the neutrino luminosities and average energies for all species, sampled in the co-moving frame of reference at a distance of about 500~km, comparing the reference simulation with muonic weak processes (blue lines) and without (red lines). Note that in the latter case, $\nu_\mu\equiv\nu_\tau$ and $\bar\nu_\mu\equiv\bar\nu_\tau$.}
\label{fig:neutrino_muonization}
\end{figure*}
\begin{figure*}[htp]
\subfigure[~At 5~ms post bounce]{
\includegraphics[width=1.0\columnwidth]{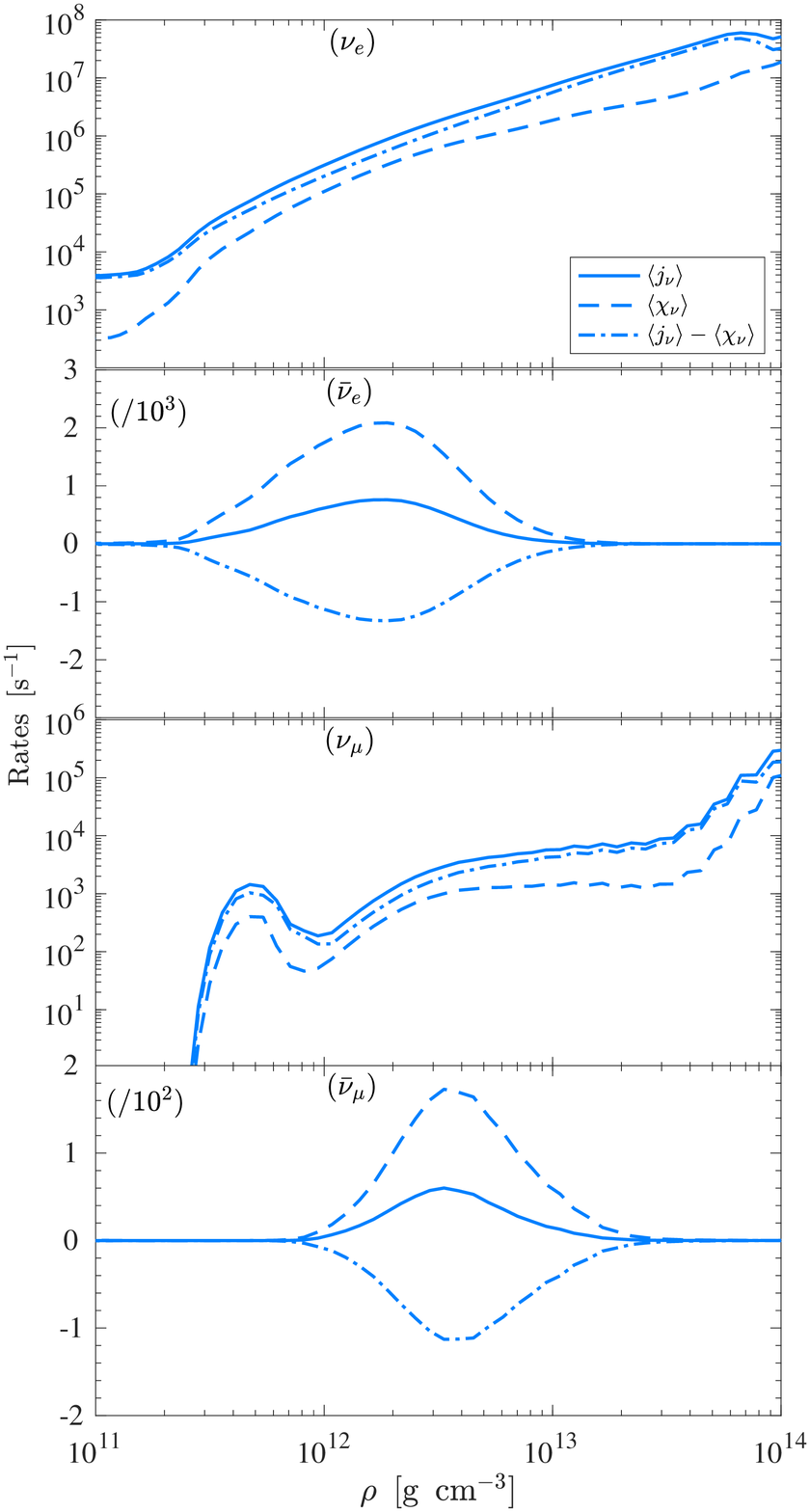}
\label{fig:rates_muonization_a}}
\subfigure[~At 30~ms post bounce]{
\includegraphics[width=1.0\columnwidth]{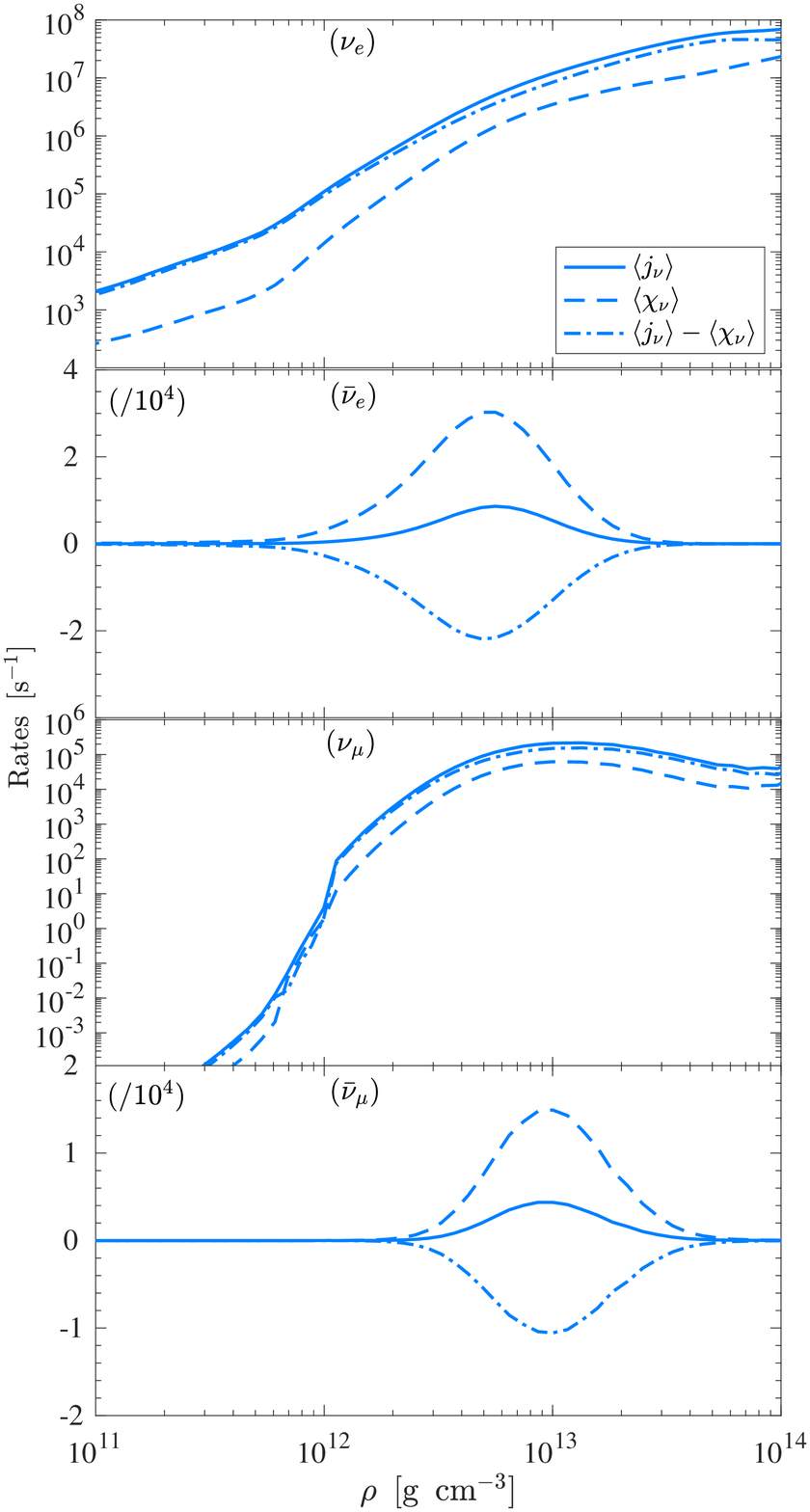}
\label{fig:rates_muonization_b}}
\caption{Density dependence of the CC rates, associated with the
neutrino emissivity  (solid line, see Eqs.~\eqref{eq:rate_j})
and the neutrino opacity (dashed line, see \eqref{eq:rate_chi})
as well as the net rate, for electron
(anti)neutrinos and muon (anti)neutrinos, which enter the
collision integral of the Boltzmann transport equation,
corresponding to the thermodynamic state at two selected post bounce
times, as shown in Figs.~\ref{fig:hydro-post-bounce_a} and
\ref{fig:hydro-post-bounce_b}.}
\label{fig:rates_muonization}
\end{figure*}

Note that the spectra of $\nu_\mu$ and $\bar\nu_\mu$ are
thermal, roughly matching the corresponding temperature profile.
Consequently the muon abundance, and the muon chemical potential
accordingly, follow the same temperature profile even though
their leading production processes~(1a) and (1b) in
Table~\ref{tab:muonic-weak} have no purely thermal character,
unlike neutrino-pair production from $e^-$--$e^+$ annihilation.
It is important to notice that the muonization is a dynamical
process. It is determined by the muonic weak rates and the
thermodynamic conditions obtained in the PNS interior. Muonic
weak equilibrium is not established instantaneously: the muon
chemical potential is significantly lower, $\mu_\mu\simeq40-90$~MeV,
than that of the electrons, $\mu_e\simeq100-200$~MeV 
(see Figs.~\ref{fig:hydro-collapse} and \ref{fig:hydro-bounce})
This situation remains during the entire post-bounce evolution,
as illustrated in Figs.~\ref{fig:hydro-post-bounce_a} and
\ref{fig:hydro-post-bounce_b} for selected times corresponding
to the early post-bounce phase. Furthermore, the temperature
profile is non-monotonic, which is well known due to the fact
that the bounce shock forms at a radius of about 10~km. The
highest temperature increase as well as the maximum temperature
obtained during the post-bounce evolution is not at the very
center of the forming PNS. Hence, the thermal production of
muon-(anti)neutrinos from pair processes results in high
average energies corresponding to the maximum temperatures
(see the bottom panel in Fig.~\ref{fig:hydro-post-bounce_b}),
which in turn gives rise to a high and continuously rising
muon abundance off center (see the top panels in
Figs.~\ref{fig:hydro-post-bounce_a} and \ref{fig:hydro-post-bounce_b})
during the later post-bounce evolution. The presence of a
finite abundance of muons at the PNS center, ranging from $Y_\mu\simeq 10^{-4}$
to a few times $10^{-3}$ corresponding to densities in excess
of few times $10^{12}$~g~cm$^{-3}$, results in substantially
higher muon-neutrino abundances than muon-antineutrinos also
during the entire post-bounce evolution (see the bottom pannels in
Figs.~\ref{fig:hydro-post-bounce_a} and \ref{fig:hydro-post-bounce_b}),
which is not observed in the reference simulation without muonic
charged current processes. Consequently, also the average energy of
$\nu_\mu$ is substantuially higher then for $\bar\nu_\mu$
in that domain due to the presence of a finite
muon chemical potential
(see Fig.~\ref{fig:hydro-post-bounce_b}), similarly as for electrons
and electron neutrinos. However, the overall
post-bounce evolution, in terms of the gross hydrodynamics
evolution, is not affected by the presence of muons and
associated muonic weak reactions after shock break out. 

It is important to emphasise here the importance of the
inelastic CC rates; i.e., with their full kinematics
implementation. A test simulation, in which we used the
elastic CC rates instead (not shown here for simplicity)
gave rise to a substantially lower muon abundance, by
nearly a factor of 2, in the off-center muon production
region associated with the highest temperatures.

\subsection{Launch of the muon neutrino burst}
\label{sec:burst}
The continuously rising muon abundance, in turn, enables the
release of a $\nu_\mu$ burst (blue solid line in the middle
panel of Fig.~\ref{fig:neutrino_muonization}), relative to 
the simulation without muonic weak processes (red curve),
associated with the shock break out. Similar to the $\nu_e$
deleptonization burst (solid curve in the top panel of
Fig.~\ref{fig:lumin_muonization}), when the shock wave crosses
the muonic neutrinosphere, muon captures on protons are enabled,
$\mu^- + p \rightarrow n + \nu_\mu$,  due to the escape of the
muon-neutrinos produced. It results not only in the substantial
rise of the muon-neutrino luminosity, relative to the case
without muonic weak processes, but is also associated with the
continuous rise of the off-central muon abundance.
The later increase by a factor of about 10 during the first
10--30~ms after core bounce (see Fig.~\ref{fig:hydro-post-bounce}). 
However, the magnitude of the
associated luminosity of the $\nu_\mu$ burst is lower by a factor
of more than five than that of the $\nu_e$ burst (see
Fig.~\ref{fig:lumin_muonization}), due to the generally lower muon abundance 
(see Fig.~\ref{fig:hydro-post-bounce}), which is related to the
slower CC rates for $\nu_\mu$ than for $\nu_e$. The corresponding 
integrated CC rates, defined as follows,
\begin{eqnarray}
&& \langle j_\nu \rangle = \frac{2\pi c}{(hc)^3}\frac{m_{\rm B}}{\rho}
\int d(\cos\vartheta) dE_\nu E_\nu^2j_\nu(E_\nu)\left[1-f_\nu(\cos\vartheta,E_\nu)\right],\nonumber\\
&&
\label{eq:rate_j}\\
&& \langle \chi_\nu \rangle = \frac{2\pi c}{(hc)^3}\frac{m_{\rm B}}{\rho}
\int d(\cos\vartheta)dE_\nu E_\nu^2\vert\chi_\nu(E_\nu)\vert \,f_\nu(\cos\vartheta,E_\nu),\nonumber \\
&&
\label{eq:rate_chi}
\end{eqnarray}
are illustrated in Fig.~\ref{fig:rates_muonization}.
The conditions in Fig.~\ref{fig:rates_muonization}
corresponding to the two situations illustrated in
Figs.~\ref{fig:hydro-post-bounce_a} at 5~ms post bounce
and \ref{fig:hydro-post-bounce_b} at 30~ms post bounce.
Note further, that the NLS as well as LFE and LFC rates are
omitted in Fig.~\ref{fig:rates_muonization}, which are
negligible compared to the CC rates.

Furthermore, the $\bar\nu_\mu$ luminosity is also affected by a
finite net muon abundance and associated muonic weak processes.
However, while the $\nu_\mu$ experience a sudden rise, as discussed
above, the $\bar\nu_\mu$ luminosity is reduced (see the blue
dash-dotted curves in the middle panel of
Fig.~\ref{fig:lumin_muonization}) relative to the simulation
without muonic weak processes (red dash-dotted curve in the
middle panel of Fig.~\ref{fig:lumin_muonization}).
Here, for $\bar\nu_\mu$, the net charged-current rates are
dominated by $\bar\nu_\mu$ absorption on neutrons, and hence
the expression~\eqref{eq:dfvebdt} is overall negative at densities
below $10^{12}$~g~cm$^{-3}$, as shown in the bottom panels of
Fig.~\ref{fig:rates_muonization}. This gives rise to anti-muon
production, in contrast to \eqref{eq:dfvedt}, which is overall
positive, acting mostly as a muon sink (see
Fig.~\ref{fig:rates_muonization})

As aforementioned, the fully inelastic muonic CC rates result
in a substantially higher muon abundance than when the elastic
CC rates are employed. Consequently, the magnitude of the luminosity
of the $\nu_\mu$ burst is somewhat higher for the fully inelastic
rates. The same holds for the magnitude of the reduced
$\bar\nu_\mu$ luminosity. Therefore, it is important to implement
the muonic CC rates in their full-kinematics treatment.

In the region of $\nu_\mu$ losses during the release of the 
$\nu_\mu$ burst, corresponding to densities between 
$\rho=10^{11}-10^{13}$~g~cm$^{-3}$; i.e., the location of the 
$\nu_\mu$ neutrinosphere, there is a slight feedback on the 
PNS structure resulting in slightly lower temperatures (see 
Fig.~\ref{fig:hydro-post-bounce_b}) compared to the simulation 
without muons. These lower temperatures affect also the 
electron (anti)neutrino luminosities and average energies;
however, only marginally (see the top panels in 
Fig.~\ref{fig:neutrino_muonization}). Moreover, the 
$\tau$-(anti)neutrino luminosities and average energies are 
affected from the slightly higher compactness achieved due 
to the additional losses associated with the $\nu_\mu$ burst.
Related is the presence of muons at the highest densities,
which  results in a slight temperature increase (see 
Fig.~\ref{fig:hydro-post-bounce_b}). This implies a 
softening of the high-density EOS, since muons are 
significantly more massive than electrons, and electrons 
are effectively replaced by muons. This feeds back partly 
to higher average energies of $\nu_{\tau}$ and $\bar\nu_\tau$,
which are produced thermally from pair processes, at 
the highest densities in the PNS interior trapping regime
(see the bottom panel in  Fig.~\ref{fig:hydro-post-bounce_b}). 

Note further that after about 50~ms post bounce the magnitude
of the $\nu_\mu$ and $\bar\nu_\mu$ luminosities will have settled 
back to about $3.0-3.5\times 10^{52}$~erg~s$^{-1}$ that 
corresponds to the value without muonic reactions. The later 
post-bounce evolution with the influence of muons and associated 
muonic weak processes has been discussed in Ref.~\cite{Bollig:2017}, 
where the potential role with respect to neutrino heating and 
cooling contributions as well as on the revival of the stalled 
bounce shock was explored.

\section{Role of convection}
\label{sec:convection}
In order to study the potential role of convection induced due
to  the presence of negative lepton number gradients, it has 
been convenient to estimate the Brunt-V{\"a}is{\"a}l{\"a} 
frequency~\cite{Wilson:1988,Keil:1996ab,Buras:2005rp,Kotake:2005zn,Roberts:2012c,Mirizzi:2016}
\begin{equation}
\omega_{\rm BV} = \rm sign\left\{C_{\rm Ledoux}\right\} \sqrt{g\,\rho^{-1}\left\vert C_{\rm Ledoux}\right\vert}~,
\label{eq:wBV}
\end{equation}
with gravitational acceleration, $g$, restmass density,
$\rho$, and the Ledoux-convection criterion $C_{\rm Ledoux}$.
The latter can be related to derivatives of thermodynamics quantities
as follows,
\begin{equation}
C_{\rm Ledoux} = \left(\frac{\partial P}{\partial s}\right)_{\rho,Y_{\rm L}}\frac{\rm d s}{\rm d r} + \left(\frac{\partial P}{\partial Y_{\rm L}}\right)_{\rho,s}\frac{\rm d Y_{\rm L}}{\rm d r}~.
\label{eq:Cledoux}
\end{equation}
The thermodynamic derivatives of the pressure, $P$, are evaluated
at constant restmass density, $\rho$, and constant lepton number,
$Y_{\rm L}$, in the case of the entropy derivative, and constant
entropy, $s$, in the case of the lepton-number derivative. Then,
convective instability is inferred when $\omega_{\rm BV}>0$. Note
that for the thermodynamic derivative, $(\partial P/\partial 
Y_{\rm L})_{\rho,s}$, finite differencing is employed based on 
the tabulated EOS, while the lepton-number gradient, 
$\rm d Y_{\rm L}/\rm d r$, is obtained by finite differencing 
of the SN simulation data.

According to the standard model each lepton number is conserved among
its flavor. The situation of lepton-number violating processes, which 
belong to the physics beyond the standard model, are not considered here.
In the presence of more than one conserved and non-zero lepton number,
$Y_{\rm L_e}$ and $Y_{\rm L_\mu}$, the total pressure \eqref{eq:P} can
be rewritten as the sum of all partial pressures as follows,
$P=P_{\rm B} + P_{\rm L_e} + P_{\rm L_\mu}$.
Consequently, both lepton numbers appear as explicit dependencies of
the density, $\rho=\rho(P,s,Y_{\rm L_e},Y_{\rm L_\mu})$, which in
turn modifies the Ledoux criterion as follows,
\begin{eqnarray}
C_{\rm Ledoux} &=&
\left(\frac{\partial P}{\partial s}\right)_{\rho,Y_{\rm L_e},Y_{\rm L_\mu}}\frac{\rm d s}{\rm d r}
+
\left(\frac{\partial P}{\partial Y_{\rm L_e}}\right)_{\rho,s,Y_{\rm L_\mu}}\frac{\rm d Y_{\rm L_e}}{\rm d r}
\nonumber \\
&&\,
+
\left(\frac{\partial P}{\partial Y_{\rm L_\mu}}\right)_{\rho,s,Y_{\rm L_e}}\frac{\rm d Y_{\rm L_\mu}}{\rm d r}
\label{eq:Cledoux_mod}
\end{eqnarray}

\begin{figure}[t!]
\includegraphics[width=0.99\columnwidth]{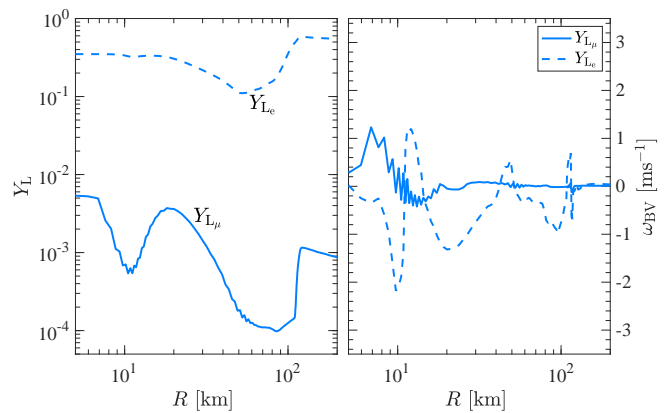}
\caption{Lepton number (left panel) and lepton number gradient (right panel) for both lepton flavors, muon (solid lines) and electron (dashed lines), at about 30~ms post bounce (see Fig.~\ref{fig:hydro-post-bounce_b}).}
\label{fig:gradient}
\end{figure}

Since the present article's concern is the impact of muons, and 
associated muonic weak processes, on the SN dynamics, the focus 
is on the lepton-number, $Y_{\rm L}$, and the associated second 
term in Eq.~\eqref{eq:Cledoux}; in particular, since it has been 
shown that the presence of muons has a negligible impact on the 
PNS structure and the entropy profile. Due to the separation of 
muonic and electronic lepton numbers, henceforth denoted as 
$Y_{\rm L_\mu}$ and $Y_{\rm L_e}$, here the following question 
shall be addressed: Can a negative muonic lepton number gradient 
drive convection? Fig.~\ref{fig:gradient} shows the lepton 
numbers (left panel) and the lepton-number gradient terms of 
$\omega_{\rm BV}$ (right panel), shortly after core bounce. 
Note that, in the case without muons and associated weak 
reactions that give rise to a finite muon abundance, the muonic 
lepton number is given by $Y_{\nu_\mu}-Y_{\bar\nu_\mu}$ and, 
hence, suppressed by several orders of magnitude, such that 
its gradient is effectively zero. In contrast, here one can 
already identify shortly after core bounce the presence of the 
additional and non-negligible muon lepton number and associated 
$\omega_{\rm BV}$ contributions (solid lines in 
Fig.~\ref{fig:gradient}). The region with $\omega_{\rm BV}>0$ 
for $Y_{\rm L_\mu}$ corresponds to the PNS interior from 
intermediate to highest densities, on the order of 
$\rho=10^{12}$~g~cm$^{-3}$ to few times $10^{14}$~g~cm$^{-3}$ 
(see Fig.~\ref{fig:hydro-post-bounce_b}), unlike for 
$Y_{\rm L_e}$ for which $\omega_{\rm BV}>0$ at lower 
densities, at the PNS surface. The region with $\omega_{\rm BV} > 0$
for $\rm Y_{\rm L_e}$ at large radii, between the SN shock and the
PNS surface, is relevant for the development of convection. 
This is essential to the neutrino heating and cooling of matter
in this region. The region with $\omega_{\rm BV} > 0$ for 
$\rm Y_{\rm L_\mu}$ indicates the occurance of convection
 in the PNS interior. It is interesting to note that the
 Ledoux-convection criterion has the same magnitude for
 $\rm Y_{\rm L_e}$  and $\rm Y_{\rm L_\mu}$ (see Fig.~\ref{fig:gradient}).
 This remains true during the 
entire post-bounce evolution. The magnitude of the impact 
remains to be determined in detailed multi-dimensional studies, 
preferably in three spatial dimensional simulations. This 
might have interesting implications for the emission of 
gravitational waves stemming from high densities \cite{Mueller:2017,Kotake:2018b,Burrows:2018b,Mueller:2019,Burrows:2019,Mezzacappa:2020}.

%

\section{Summary}
\label{sec:summary}
In the present article, the extension of {\tt AGILE-BOLTZTRAN} to
treat 6-species neutrino transport was introduced, overcoming 
the previous drawback of assuming equal $\mu$-- and 
$\tau$--(anti)neutrino distributions. This enables a variety of 
muonic weak processes. These include muonic charged-current 
emission/absorption reactions involving the mean-field nucleons, 
as well as purely leptonic flavor-changing reactions. All these 
muonic rates have significant inelastic contributions. Hence, it 
is important to treat the corresponding phase space properly. 
In addition to the disentanglement of $\mu$-- and 
$\tau$--(anti)neutrinos in the transport scheme, the presence 
of muonic weak processes gives rise to a finite and rising 
muon abundance. Therefore, together with its corresponding 
evolution equation, the muon abundance was added as an 
additional independent variable to the neutrino 
radiation-hydrodynamics state-vector of {\tt AGILE-BOLTZTRAN}.

With these updates, stellar core collapse was simulated and 
studied in detail with a particular focus on the appearance of 
muons, muonic weak reactions, and possible consequences 
for SN phenomenology. It had been claimed previously that 
muons and their associated weak reactions may enhance 
the neutrino heating efficiency in multidimensional SN 
simulations under certain circumstances during the 
post-bounce evolution~\cite{Bollig:2017}. Beginning with 
our first focus, muonization of SN matter, we find that it 
starts shortly before core bounce in two steps: {\em first}, 
from the production of high-energy muon-(anti)neutrinos 
from neutrino-pair processes, and {\em second}, from the 
absorption of these high-energy muon-neutrinos as part of 
the muonic charged-current and lepton-flavor changing 
processes. Here we conclude that the importance of the 
charged-current reactions exceed the latter by far for the 
muonization. We find that the muon abundance rises by 
more than one order of magnitude during the first few ten 
milliseconds post bounce, reaching values of 
$Y_\mu\simeq 10^{-4}$ to a few times $10^{-3}$ -- in 
particular, off center associated with the increasing 
temperature there. With regard to SN phenomenology: 
The presence of muons leads to a muon-neutrino burst shortly 
after core bounce, which is due to the shock propagation 
across the muon-neutrinosphere, similar to what happens 
with regard to the electron-neutrino burst when the shock 
passes the electron neutrinosphere. However, the 
muon-neutrino burst has a lower magnitude due to the 
significantly lower muon abundance and muonic 
charged-current weak rates, relative to the abundances 
and rates associated with electrons. With  regard to the 
evolution of the PNS: It is interesting to note that muons 
and muon-(anti)neutrinos are not in weak equilibrium 
instantaneously, unlike the electrons and positrons with 
their neutrino species. That is, the electron chemical 
potentials by far exceed the muon chemical potentials 
at the PNS interior. Only towards low densities near 
the neutrinospheres at the PNS surface, where the 
abundance of trapped neutrinos drops to zero, are the electron 
and muon chemical potentials equal during the 
post-bounce evolution. It remains to be explored in future 
studies how the muons approach equilibrium during the 
later PNS deleptonization phase; i.e., after the onset  of 
the SN explosion on a timescale of several ten seconds.
Furthermore, the presence of an additional muon 
lepton-number gradient may impact convection at high 
densities in the PNS interior. To confirm this requires 
multidimensional simulations, which cannot be investigated 
here. We find that the presence of a finite and continuously 
rising muon abundance in the PNS interior has a softening 
impact on the high-density equation of state. This, in turn, 
is known to result in smaller PNS and shock radii during 
the long-term post-bounce supernova evolution on the 
order of several hundreds of milliseconds, confirming the
findings of Ref.~\cite{Bollig:2017}, which has the potential 
of enhancing the neutrino heating efficiency through higher 
neutrino energies and luminosities. The latter is also known 
from multidimensional simulations comparing stiff and 
soft hadronic equations of state \cite{Suwa:2013}. 
Moreover, the finite muon abundance may also give rise 
to weak processes involving pions \cite{Reddy:2020}, which 
remains to be explored in future studies.

\begin{figure*}[t!]
\centering
\subfigure[~Lumonisty]{
\includegraphics[width=0.975\columnwidth]{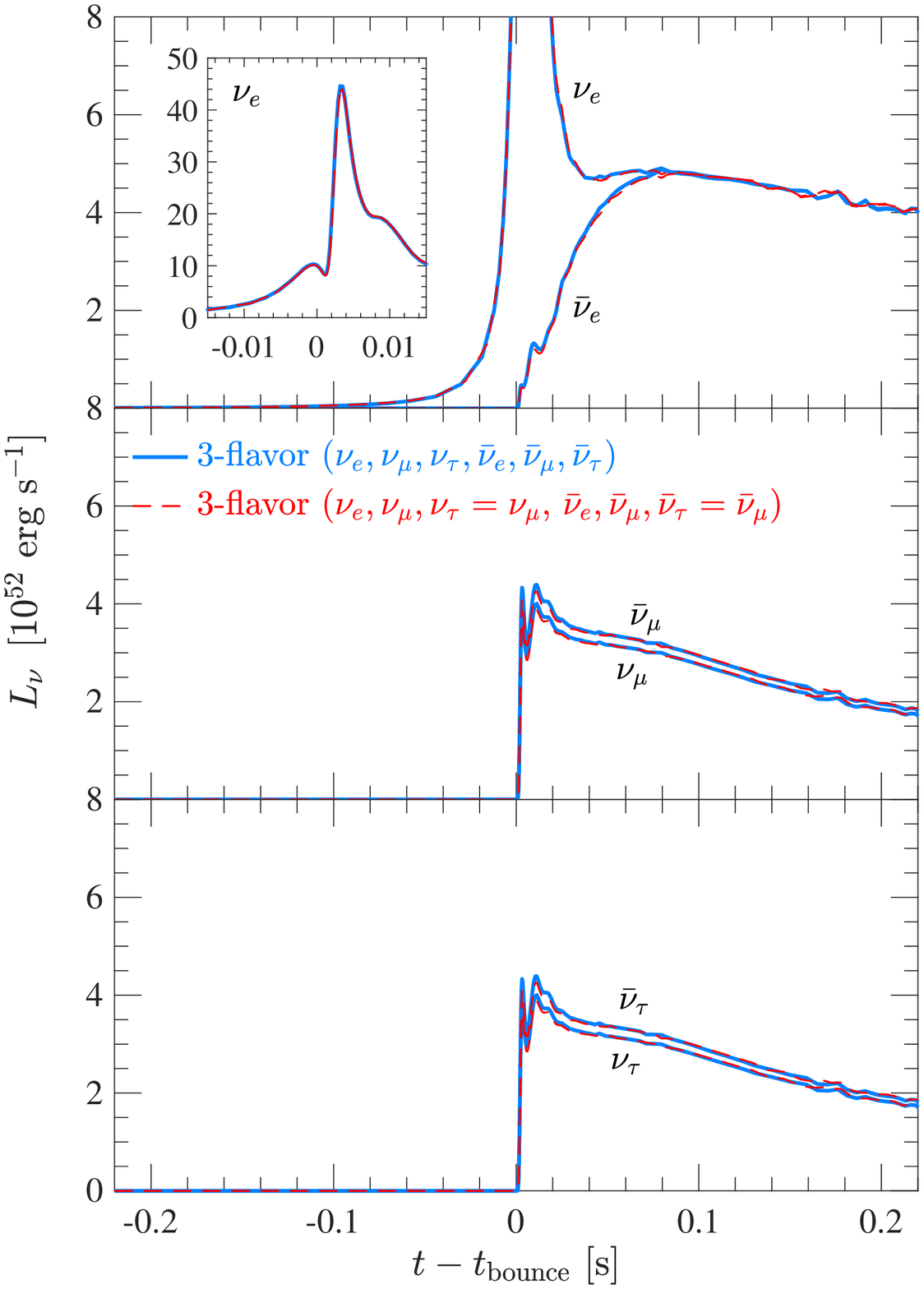}\label{fig:lumin_appendix}
}
\hfill
\subfigure[~Mean energy]{
\includegraphics[width=0.96\columnwidth]{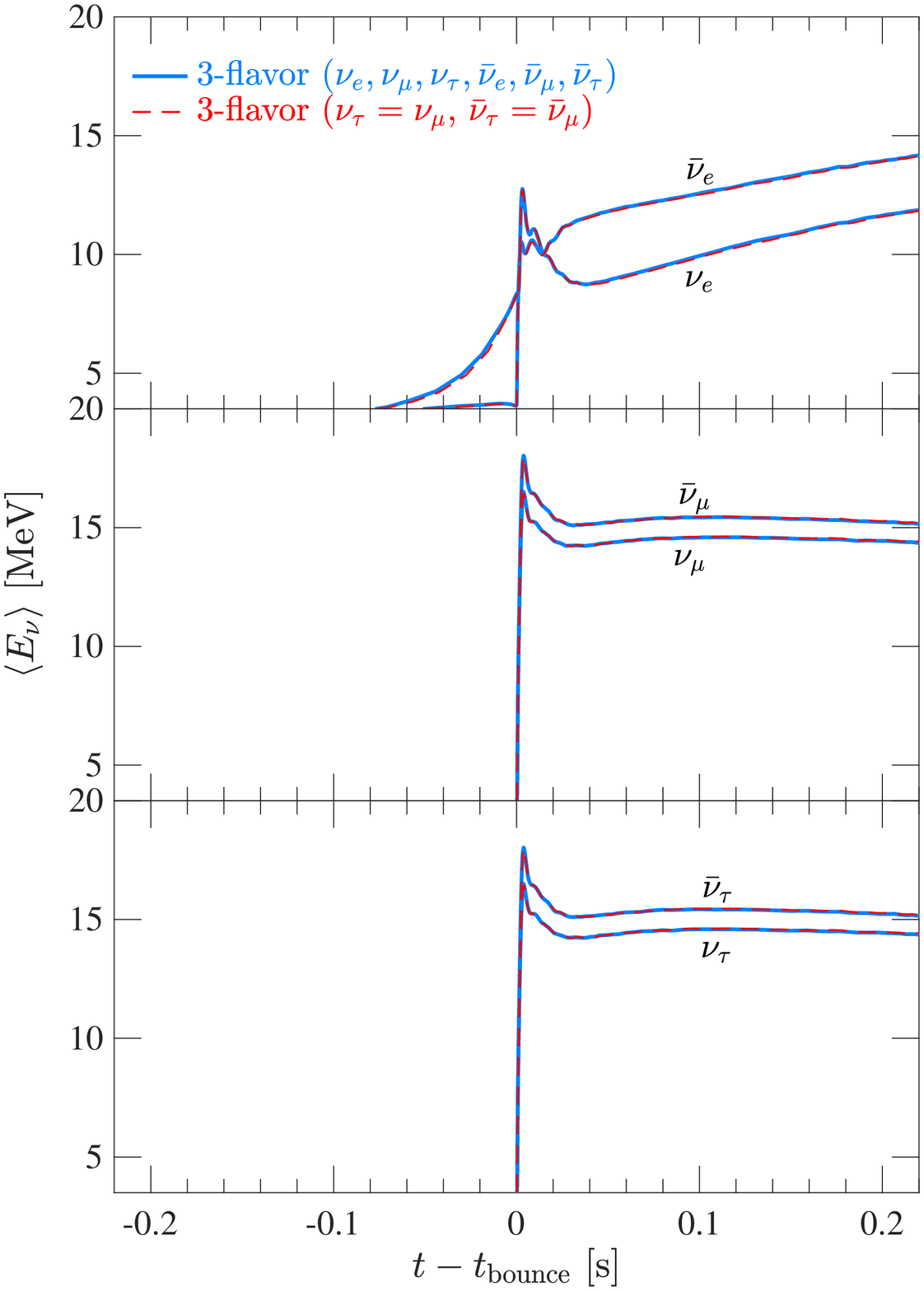}\label{fig:rms_appendix}
}
\caption{(Color online) Neutrino luminosities and average energies, sampled in the co-moving frame of reference, comparing the reference run employing the {\em actual} 3-flavor neutrino transport scheme (blue lines) and {\em mimicking} 3-flavor (red lines).}
\label{fig:neutrino_appendix}
\end{figure*}

\begin{acknowledgments}
T.F. acknowledges support from the Polish National Science Center (NCN) under grant No. 2016/23/B/ST2/00720 and No 2019/33/B/ST9/03059. G.M.P is partly supported by the Deutsche Forschungsgemeinschaft (DFG, German Research Foundation)---Project-ID 279384907--- SFB~1245. A.M. acknowledges support from the National Science Foundation through grant NSF PHY~1806692. The SN simulations were performed at the Wroclaw Center for Scientific Computing and Networking (WCSS) in Wroclaw (Poland).
\end{acknowledgments}

\appendix
\section{{\tt AGILE-BOLTZTRAN} -- Extension to 6-species Boltzmann neutrino transport}
\label{sec:A} 
Here, a comparison is presented between the reference run employing the 6-species neutrino transport scheme (blue lines) and the run based on the traditional 4-species (red lines), where besides the inclusion of electron and muon neutrino flavors it is assumed that the tau-(anti)neutrino distributions are equal to the muon-(anti)neutrino distributions. No muonic weak processes are considered here. All muonic weak rates are set (numerically) to zero.

Fig.~\ref{fig:neutrino_appendix} compares the evolution of the neutrino luminosities in Fig.~\ref{fig:lumin_appendix} and the average neutrino energies in Fig.~\ref{fig:rms_appendix} for the two SN simulations, respectively. Otherwise, both simulations employ an identical set of input physics, as introduced in Secs.~\ref{sec:SNmodel_EOS} and \ref{sec:SNmodel_transport}. The relative change between the two runs in terms of neutrino losses is on the order of less than a few tenths of one percent, attributed mostly to a slightly different converged solution of the radiation-hydrodynamics equations~\cite{Liebendoerfer:2004} with the implementation of the muon abundance as additional independent degree of freedom. Furthermore, the entire SN hydrodynamics -- i.e., shock formation, shock evolution, and post-bounce mass accretion -- shows no quantitative differences.

\section{Implementation of muonic weak processes}
\label{sec:B}

\subsection{Muonic charged-current processes -- elastic rates}
\label{sec:B1}
The CC rates within the full kinematics approach, including self-consistent contributions from weak magnetism, are provided in Ref.~\cite{Fischer:2020} for the electronic CC processes. They have been reviewed recently for the muonic reactions~(1a) and (1b) of Table~\ref{tab:muonic-weak} in Ref.~\cite{Guo:2020}. For the comparison with this full kinematics treatment, it is convenient to provide CC muonic rates in the elastic approximation; i.e., assuming a zero-momentum transfer, for which the absorptivity ($\nu_\mu+n\rightarrow p + \mu^-$) is given by the following analytical expression:
\begin{widetext}

\begin{eqnarray}
&&
\chi_{\nu_\mu}(E_{\nu_\mu})  = \frac{G_{\rm F}^2}{\pi} \left(g_V^2 + 3 g_A^2\right) \, E_\mu^2 \, \sqrt{1-\left(\frac{m_\mu}{E_\mu}\right)^2} \, \left[1-f_\mu(E_\mu)\right] \frac{n_n-n_p}{1-\exp\left\{\frac{\varphi_p-\varphi_n}{T}\right\}}~,
\label{eq:cc_rate}
\end{eqnarray}
\end{widetext}

with Fermi constant, $G_{\rm F}$, vector and axial-vector coupling constants, $g_V=1.0$ and $g_A=1.27$, as well as with equilibrium Fermi--Dirac distribution functions for the muons, $f_\mu(E_\mu;\{\mu_\mu,T\})$, neutron/proton number densities, and the latter free Fermi gas chemical potentials, $n_{n/p}$ and $\varphi_{n/p}$, respectively. The latter are related to the nuclear EOS chemical potentials, $\mu_i$, as follows: $\varphi_{n/p} = \mu_{n/p} - m_{n/p}^* - U_{n/p}$, with neutron/proton single-particle vector-interaction potentials $U_{n/p}$ and effective masses $m_{n/p}^*$ \cite{MartinezPinedo:2012,Roberts:2012a}, both of which are given by the nuclear EOS. Inelastic contributions and weak-magnetism corrections are approximately taken into account via neutrino-energy-dependent multiplicative factors to the emissivity and opacity~\cite{Horowitz:2001xf}. A similar expression as \eqref{eq:cc_rate} is obtained for $\bar\nu_\mu$ by replacing the muon Fermi distribution with that of anti-muons and replacing $n\leftrightarrow p$ for the neutron/proton number densities and free gas chemical potentials.

The emissivity and absorptivity are related intimately via detailed balance,
\begin{equation}
j_{\nu_\mu}(E_{\nu_\mu}) = \exp\left\{-\frac{E_{\nu_\mu}-\mu_{\nu_\mu}^{\rm eq}}{T}\right\}\,\chi_{\nu_\mu}(E_{\nu_\mu})~,
\end{equation}
with muon-neutrino equilibrium chemical potential given by the expression, $\mu_{\nu_\mu}^{\rm eq} = \mu_\mu - \left(\mu_n-\mu_p\right)$.

For this limited kinematics assuming zero-momentum transfer, one can relate the muon and $\nu_\mu$ energies as follows,
\begin{equation}
E_{\mu^\pm} = E_{(\bar\nu_\mu)\nu_\mu}\mp(m_n-m_p)\mp(U_n-U_p)~,
\label{eq:EvEmu}
\end{equation}
with the medium-modified $Q$ value, 
$m_\mu\pm(m_n-m_p)\pm(U_n-U_p)$~\cite{Reddy:1998,MartinezPinedo:2012,Roberts:2012a}.
Note that, as for the electron-flavor neutrinos, the nuclear medium modifications for the charged-current 
rate at the mean-field level modify the opacity substantially with increasing density~\cite{MartinezPinedo:2014}. In particular, at densities in excess of 
$\rho=10^{13}$~g~cm$^{-3}$, where muons can be expected, the opacity drop can 
differ significantly from the vacuum $Q$ value, $m_\mu\pm(m_n-m_p)$ due to the large 
difference of the single particle potentials can well be on the order of $U_n-U_p=40-80$~MeV, 
depending on the nuclear EOS~\cite{Fischer:2014,MartinezPinedo:2014}.

The presence of high-energy $\nu_\mu$ and $\bar\nu_\mu$ enables the production of $\mu^\pm$. The collision integrals for $\nu_\mu$ and $\bar\nu_\mu$ for the reactions (1a) and (1b) of Table~\ref{tab:muonic-weak} take the following form:
\begin{eqnarray}
\left.\frac{\partial F_{\nu_\mu}}{c\,\partial t}(E_{\nu_\mu},\vartheta)\right\vert_{\rm CC}
&=& 
\frac{j_{\nu_\mu}(E_{\nu_\mu})}{\rho} - \tilde\chi_{\nu_\mu}(E_{\nu_\mu})F_{\nu_\mu}(E_{\nu_\mu},\vartheta) \nonumber \\
&& \label{eq:dfvedt} \\
\left.\frac{\partial F_{\bar\nu_\mu}}{c\,\partial t}(E_{\bar\nu_\mu},\vartheta)\right\vert_{\rm CC}
&=& 
\frac{j_{\bar\nu_\mu}(E_{\bar\nu_\mu})}{\rho} - \tilde \chi_{\bar\nu_\mu}(E_{\bar\nu_\mu})F_{\bar\nu_\mu}(E_{\bar\nu_\mu},\vartheta)~,\nonumber \\
&& \label{eq:dfvebdt}
\end{eqnarray}
with effective opacity defined as follows, $\tilde\chi_\nu=\chi_\nu+j_\nu$~\cite{Mezzacappa:1993gn,Mezzacappa:1993gm}. Expressions~\eqref{eq:dfvedt} and \eqref{eq:dfvebdt} are equivalent to those for the electron-(anti)neutrinos with electronic charged-current emissivity and opacity~\cite{Fischer:2012a}. The muon abundance, $Y_\mu$, is then added as an independent variable to the {\tt AGILE} state vector, for which the following differential-integral evolution equation is solved,

\begin{widetext}

\begin{eqnarray}
\label{eq:ymueqn}
&&
\left.\frac{\partial Y_\mu}{\partial t}\right\vert_{\rm CC}
= 
\left.\frac{2\pi m_{\rm B}}{\left(hc\right)^3}
\right[
\int dE_{\bar\nu_\mu} dE_{\bar\nu_\mu}^2 d\left(\cos\vartheta\right)\left.\frac{\partial F_{\bar\nu_\mu}}{\partial t}(E_{\bar\nu_\mu},\vartheta)\right\vert_{\rm CC}
-
\left.
\int dE_{\nu_\mu} dE_{\nu_\mu}^2 d\left(\cos\vartheta\right)
\left.\frac{\partial F_{\nu_\mu}}{\partial t}(E_{\nu_\mu},\vartheta)\right\vert_{\rm CC}
\right]~,
\end{eqnarray}
with baryon mass $m_{\rm B}=938$~MeV. Equation~\eqref{eq:ymueqn} is similar to the evolution equation for $Y_e$ (see Eqs.~(17)--(25) in Ref.~\cite{Mezzacappa:1993gm}), where, instead, the electronic charged-current neutrino emissivity and opacity are used (see expressions~(6) and (7a)--(7d) in Ref.~\cite{Fischer:2012a}). 

\subsection{Neutrino-muon scattering}
\label{sec:B2}
For neutrino--lepton scattering (NLS), $\nu + l^\pm \leftrightarrows l'^\pm + \nu'$, distinguishing here between neutrinos, $\nu\in\{\nu_e,\nu_\mu,\nu_\tau\}$, and leptons, $l^\pm\in\{e^\pm,\mu^\pm,\tau^\pm\}$, the collision integral is given by the following integral expression,
\begin{eqnarray}
\left.\frac{\partial F_{\nu}}{c\,\partial t}\right\vert_{\rm NLS}(E_{\nu},\vartheta) 
&=&
\left[\frac{1}{\rho}-F_{\nu}(E_{\nu},\vartheta)\right]
\frac{1}{(hc)^3} \frac{1}{c} 
\int E_{\nu'}^2 dE_{\nu'} \int d\left(\cos\vartheta'\right) \int d\phi \,
\mathcal{R}^{\rm in}_{\rm NLS, \nu}(E_{\nu},E_{\nu'},\cos\theta) \, F_{\nu'}(E_{\nu'},\vartheta')
\nonumber
\\
&&\;-
F_{\nu}(E_{\nu},\vartheta)\frac{1}{(hc)^3} \frac{1}{c} \int E_{\nu'}^2 dE_{\nu'} \int d\left(\cos\vartheta'\right) \int d\phi \,
\mathcal{R}^{\rm out}_{\rm NLS, \nu}(E_{\nu},E_{\nu'},\cos\theta) \left[\frac{1}{\rho}-F_{\nu'}(E_{\nu'},\vartheta')\right]
\label{eq:NMS_collision_integral}
\end{eqnarray}
with the following definition for the in- and out-scattering kernels,
\begin{eqnarray}
&&
\mathcal{R}^{\rm in}_{\rm NLS, \nu}(E_{\nu},E_{\nu'},\cos\theta) 
=
\int\frac{d^3 p_{l}}{(2\pi\hbar c)^3}\frac{d^3 p_{l'}}{(2\pi\hbar c)^3} 
2\,f_{l'}(E_{l'})\left[1-f_{l}(E_{l})\right] 
\frac{\sum_{\rm s} \left\vert\mathcal{M}\right\vert^2_{\nu + l \leftarrow l' + \nu'}}{16 E_{\nu} E_{l} E_{\nu'} E_{l'}} \,
(2\pi)^4\delta^4(p_{\nu}+p_{l}-p_{l'}-p_{\nu'})~, \;\;\;
\label{eq:R1nms}
\\
&&
\mathcal{R}^{\rm out}_{\rm NLS, \nu}(E_{\nu},E_{\nu'},\cos\theta) 
=
\int\frac{d^3 p_{l}}{(2\pi\hbar c)^3}\frac{d^3 p_{l'}}{(2\pi\hbar c)^3} 
2 f_{l}(E_{l})\left[1-f_{l'}(E_{l'})\right] 
\frac{\sum_{\rm s} \left\vert\mathcal{M}\right\vert^2_{\nu + l \rightarrow l' + \nu'}}{16 E_{\nu} E_{l} E_{\nu'} E_{l'}} \, 
(2\pi)^4\delta^4(p_{\nu}+p_{l}-p_{l'}-p_{\nu'})~, \;\;\;
\label{eq:R2nms}
\end{eqnarray}
with equilibrium Fermi-Dirac distribution functions $f_l(E_l)$ for initial- and final-state leptons. In addition to the energy difference between incoming and outgoing neutrinos, $E_{\nu'}-E_\nu$, the scattering kernels depend on the total momentum scattering angle between the incoming and outgoing neutrino, defined as follows,
\begin{eqnarray}
\cos\theta = \cos\vartheta\cos\vartheta' - \sqrt{\left(1-\cos\vartheta\right)\left(1-\cos\vartheta\right)}\cos\phi~,
\label{eq:costheta}
\end{eqnarray}
with lateral momentum angles $(\vartheta,\vartheta')$ and relative azimuthal angle $\phi=\varphi-\varphi'$ (for illustration, see Fig.~(1) in Ref~\cite{Mezzacappa:1993gn}).
 
Here, for neutrino--muon scattering (NMS), the approach for
neutrino--electron scattering (NES) is extended following the
Refs.~\cite{Tubbs:1975jx,SchinderShapiro:1982,Bruenn:1985en,Mezzacappa:1993gx}. As
an example, in the following neutrino--muon scattering,
$\nu + \mu \leftrightarrows \mu' + \nu'$, will be considered, which
contains neutral-current $Z^0$-boson and charged-current $W^-$-boson
interactions, similar to $\nu_e-e^-$ scattering (see Fig.~1 in
Ref.~\cite{Tubbs:1975jx}). The matrix element, $\mathcal{M}$, for
$\nu_e-e^-$ scatteringa can be obtained from the literature,
cf. Eqs.~(C46)--C(48) in Ref.~\cite{Bruenn:1985en}, by replacing the
electron and $\nu_e$ spinors,
$\left(u_e(p_e), u_{\nu_e}(p_{\nu_e})\right)$, with those of the muon
and $\nu_\mu$, $\left(u_\mu(p_\mu), u_{\nu_\mu}(p_{\nu_\mu})\right)$,
respectievely,
\begin{equation}
\mathcal{M}_{\nu_\mu + \mu \rightarrow \mu' + \nu'_\mu} = \frac{G_{\rm F}}{\sqrt{2}} 
[ \bar{u}_{\nu_\mu}(p'_{\nu_{\mu}'}) \gamma^k \left(1-\gamma_5\right) u_{\nu_\mu}(p_{\nu_\mu}) ] 
[ \bar{u}_{\mu}(p'_{\mu}) \gamma_k \left(C_V-C_A\gamma_5\right) u_{\mu}(p_{\mu}) ]~,
\end{equation}
The matrix element depends on the particle's 4-momenta, $p_i$, and the dash denotes final states. The quantities $C_V$ and $C_A$ are the vector and axial-vector coupling constants. After spin-averaging and squaring, the transition amplitude takes the following form,
\begin{eqnarray}
&&
\qquad\qquad\qquad\qquad
\sum_{\rm s} \left\vert\mathcal{M}\right\vert^2_{\nu_\mu + \mu \rightarrow \mu' + \nu'_\mu}
=
\beta_1 M_1 + \beta_2 M_2 + \beta_3 M_3~\\
&&= 
16\,G_{\rm F}^2
\left\{
\beta_1 \, ( p_{\mu} \cdot p_{\nu_\mu} ) (p'_{\mu} \cdot p'_{\nu_\mu} ) +
\beta_2 \, ( p'_{\mu} \cdot p_{\nu_\mu} ) (p_{\mu} \cdot p'_{\nu_\mu} ) +
\beta_3 \, m_\mu^2 \, ( p_{\nu_\mu} \cdot p'_{\nu_\mu})
\right\}
\end{eqnarray}
with $\beta_1=(C_V + C_A)^2$, $\beta_2 =(C_V - C_A)^2$ and $\beta_3=C_A^2 - C_V^2$, where the values for $C_V$ and $C_A$ are listed in Table~\ref{tab:cvca}.

The individual kinematic integral expression, denoted as $I_1-I_3$, which correspond to the muon initial- and final-state momentum integrals of $M_1-M_3$, can be obtained from Eq.~(10b) in Ref.~\cite{Tubbs:1975jx} by replacing the electron with the muon 4-momenta. The remaining integrals are solved numerically following the approach developed in Ref.~\cite{Mezzacappa:1993gx} for neutrino-electron scattering, such that
\begin{eqnarray}
\mathcal{R}^{\rm out}_{\rm NMS, \nu_\mu}(E_{\nu_\mu},E_{\nu'_\mu},\cos\theta)
=
\frac{G_{\rm F}^2}{2\pi^2} \frac{1}{E_{\nu_\mu}E_{\nu'_\mu}}
\left\{
\beta_1 I_1(E_{\nu_\mu},E_{\nu'_\mu},\cos\theta) + \beta_2 I_2(E_{\nu_\mu},E_{\nu'_\mu},\cos\theta) + \beta_3 I_3(E_{\nu_\mu},E_{\nu'_\mu},\cos\theta)
\right\}.\qquad\;\;
\label{eq:nes_I}
\end{eqnarray}

\end{widetext}

The definition of the remaining integrals, $I_i(E_{\nu_\mu},E_{\nu'_\mu},\cos\theta)$, is given in Eqs.~(11)--(27) in Ref.~\cite{Mezzacappa:1993gx}, based on the polylogarithm functionals, which are used to perform the remaining Fermi-integrals.

In order to eliminate the remaining dependence on the relative azimuthal angle $\phi$ in the scattering kernels~\eqref{eq:R1nms} and \eqref{eq:R2nms}, a numerical 
32-point Gauss quadrature integration is employed, which is identical to the one of {\tt BOLTZTRAN} for neutrino--electron scattering (see Eq.~(32) in Ref.~\cite{Mezzacappa:1993gx}), such that the scattering kernels depend only on incoming and outgoing neutrino energies, as well as on the incoming and outgoing neutrino lateral angles, $\mathcal{R}^{\rm in/out}(E_{\nu_\mu},E_{\nu'_\mu},\vartheta,\vartheta')$. For the procedure to avoid singular forward scattering, expressions~(37)--(43) in Ref.~\cite{Mezzacappa:1993gx} are employed here for neutrino--muon scattering. 

For muon-antineutrino scattering on muons the expressions are a cross channel of muon--neutrino scattering introduced above (similar to the relationship between electron-antineutrino scattering on electrons and electron-neutrino scattering on electrons~\cite{Bruenn:1985en}), given by the substitution $p_{\nu_\mu}\leftrightarrow p'_{\nu_\mu}$ in the matrix element. It has been realized for neutrino--electron scattering, this corresponds to the replacement of $C_A\leftrightarrow -C_A$ in the expressions for the scattering kernels~\cite{Bruenn:1985en}. Therefore, it is straightforward to obtain the corresponding scattering kernel $\mathcal{R}^{\rm out}_{\rm NMS, \bar\nu_\mu}$. Similar replacements are done for electron-(anti) neutrino scattering on muons, $\mathcal{R}^{\rm out}_{\rm NMS, \nu_e(\bar\nu_e)}$~\cite{Bruenn:1985en}. Table~\ref{tab:cvca} summarizes the values of $C_V$ and $C_A$ for all neutrino-(anti)muon scattering reactions. Furthermore, since the collision integral of the Boltzmann equation~\eqref{eq:NMS_collision_integral} has an identical form for NMS and NES, the general neutrino--lepton scattering kernel in the module for the inelastic scattering processes of {\tt Boltztran} is defined as follows,
\begin{equation}
\mathcal{R}^{\rm out}_{\rm NLS,\nu} := \mathcal{R}^{\rm out}_{\rm NES, \nu} + \mathcal{R}^{\rm out}_{\rm NMS, \nu}~,
\end{equation}
for each pair of neutrino specie $\nu$.

Note that, due to detailed balance, the transition amplitudes for in- and out-scattering are equal, $\sum_{\rm s} \left\vert\mathcal{M}\right\vert^2_{\nu + l \rightarrow l' + \nu'}=\sum_{\rm s} \left\vert\mathcal{M}\right\vert^2_{\nu + l \leftarrow l' + \nu'}$ (the degeneracy factors cancel), such that the scattering kernels~\eqref{eq:R1nms} and \eqref{eq:R2nms} are related via,

\begin{widetext}

\begin{eqnarray}
&&
\mathcal{R}^{\rm in}_{\rm NLS,\nu}(E_{\nu},E_{\nu'},\vartheta,\vartheta') = \mathcal{R}^{\rm out}_{\rm NLS,\nu}(E_{\nu},E_{\nu'},\vartheta,\vartheta') \, \exp\left\{-\frac{E_\nu-E_\nu'}{T}\right\}~,
\end{eqnarray}
which is the case for both, NES and NMS.

\subsection{Purely leptonic lepton flavor changing processes}
\subsubsection{{\bf Lepton flavor exchange (LFE)}}
\label{sec:B3a}
As an example, in the following the focus will be on the the reaction: $\nu_{\mu} + e^- \leftrightarrows \mu^- + \nu_{e}$. In close analogy to \eqref{eq:NMS_collision_integral}, the collision integral of the Boltzmann transport equation is given by the following integral expression,
\begin{eqnarray}
&& 
\left.\frac{\partial F_{\nu_\mu}}{c\, \partial t}\right\vert_{\rm LFE}(E_{\nu_\mu},\vartheta) 
=
\left[\frac{1}{\rho}-F_{\nu_\mu}(E_{\nu_\mu},\vartheta)\right]
\frac{1}{(hc)^3}\frac{1}{c} \int E_{\nu_e}^2 dE_{\nu_e} \int d\left(\cos\vartheta'\right) \int d\phi \, 
\mathcal{R}^{\rm in}_{\rm LFE, \nu_\mu}(E_{\nu_\mu},E_{\nu_e},\cos\theta) \,
F_{\nu_e}(E_{\nu_e},\vartheta')
\nonumber
\\
&&\;\;\;\;\;\;\;\;\;\;\;\;\;\;
-
F_{\nu_\mu}(E_{\nu_\mu},\vartheta)
\frac{1}{(hc)^3} \frac{1}{c} \int E_{\nu_e}^2 dE_{\nu_e} \int d\left(\cos\vartheta'\right) \int d\phi \,
\mathcal{R}^{\rm out}_{\rm LFE, \nu_\mu}(E_{\nu_\mu},E_{\nu_e},\cos\theta)
\left[\frac{1}{\rho}- F_{\nu_e}(E_{\nu_e},\vartheta')\right]~, 
\label{eq:leptonic_a_dfdt}
\end{eqnarray}
with the in- and out-scattering kernels, again in close analogy to \eqref{eq:R1nms} and \eqref{eq:R2nms}, given as follows,
\begin{eqnarray}
\mathcal{R}^{\rm in}_{\rm LFE, \nu_\mu}(E_{\nu_\mu},E_{\nu_e},\cos\theta) 
&=&
\int\frac{d^3p_\mu}{(2\pi\hbar c)^3}\frac{d^3p_e}{(2\pi\hbar c)^3} \,2\, f_\mu(E_\mu)\left[1-f_e(E_e)\right] \; \times \nonumber \\
&&
\;\;\;\;\;\;\;\;\;\;\times\;
\frac{\sum_{\rm s} \left\vert\mathcal{M}\right\vert^2_{\nu_\mu + e^- \leftarrow \mu^- + \nu_e}}{16 E_{\nu_\mu} E_{e} E_{\mu} E_{\nu_e}} \, 
(2\pi)^4 \delta^4(p_{\nu_e}+p_\mu-p_e-p_{\nu_\mu})
\label{eq:Rin_leptonic_a}
\\
\mathcal{R}^{\rm out}_{\rm LFE, \nu_\mu}(E_{\nu_\mu},E_{\nu_e},\cos\theta) 
&=&
\int\frac{d^3p_\mu}{(2\pi\hbar c)^3}\frac{d^3p_e}{(2\pi\hbar c)^3} \,2\, f_e(E_e)\left[1-f_\mu(E_\mu)\right] \; \times \nonumber \\
&&
\;\;\;\;\;\;\;\;\;\;\times\;
\frac{\sum_{\rm s} \left\vert\mathcal{M}\right\vert^2_{\nu_\mu + e^- \rightarrow \mu^- + \nu_e}}{16 E_{\nu_\mu} E_{e} E_{\mu} E_{\nu_e}} \, 
(2\pi)^4 \delta^4(p_{\nu_\mu}+p_e-p_\mu-p_{\nu_e})~.
\label{eq:Rout_leptonic_a}
\end{eqnarray}
The similarity to the expressions for neutrino--muon scattering, introduced above, is striking. However, unlike $\nu_\mu$--$\mu^-$ scattering which has neutral-current $Z^0$-boson and charged-current $W^-$-boson contributions, this process is given by a $W^-$-boson exchange only, with the following matrix element:
\begin{equation}
\mathcal{M}_{\nu_\mu + e^- \rightarrow \mu^- + \nu_e} = \frac{G_{\rm F}}{\sqrt{2}} 
\left[ \bar{u}_{\nu_\mu}(p_{\nu_\mu}) \gamma^k \left(1-\gamma_5\right) u_{e}(p_{e}) \right] 
\left[ \bar{u}'_{\mu}(p_{\mu}') \gamma^k \left(1-\gamma_5\right) u_{\nu_e}'(p_{\nu_e}') \right]~,
\end{equation}
with the spinors, $u_i$, depending on the corresponding 4-momenta, $p_i$, where the dash denotes final states. After summation and spin-averaging, the transition amplitude takes the following form,
\begin{equation}
\sum_{\rm s} \left\vert\mathcal{M}\right\vert^2_{\nu_\mu + e^- \rightarrow \mu^- + \nu_e}
=
64\,G_{\rm F}^2 \left(p_{\nu_\mu}\cdot p_{e}\right) \left(p_{\mu}' \cdot p_{\nu_e}'\right)~,
\label{eq:M_nms_flavor_change}
\end{equation}
such that,
%
%
%
\begin{eqnarray}
\mathcal{R}^{\rm out}_{\rm LFE, \nu_\mu}(E_{\nu_\mu},E_{\nu_e},\cos\theta)
=
4\frac{G_{\rm F}^2}{2\pi^2} \frac{1}{E_{\nu_\mu}E_{\nu_e}}
\, I_1(E_{\nu_\mu},E_{\nu_e},\cos\theta)~,
\end{eqnarray}
with the same definition of the remaining phase-space integral $I_1(E_{\nu_\mu},E_{\nu_e},\cos\theta)$ as for the case of neutrino--muon scattering discussed above. However, due to different initial-state electron and final-state muon rest masses, there are additional terms, which scale with the rest-mass energy difference, $\triangle m_{\mu e}:= (m_\mu^2-m_e^2)\,c^4/2$. Comparing these terms with those for neutrino--muon scattering (see Eqs.~(11)--(18) in Ref.~\cite{Mezzacappa:1993gx}) and using the same nomenclature as in Ref.~\cite{Mezzacappa:1993gx}, the following modifications arise,
\begin{eqnarray}
&&
I_1(E_{\nu_\mu},E_{\nu_e},\cos\theta) 
=
\frac{2\pi\,Tf_\gamma(E_{\nu_e}-E_{\nu_\mu})}{\triangle^5} \, \times
 \nonumber \\
&&
\;\;\;\; \times \;\;
\left\{
E_{\nu_\mu}^2 E_{\nu_e}^2 (1-\cos\theta)^2
\left\{
A \,T^2 \, [G_2(y_0) + 2 y_0 G_1(y_0) + y_0^2 G_0(y_0)]
+
B \, T \, [G_1(y_0) + y_0 G_0(y_0)]
+
C \, G_0(y_0) 
\right\}
\right.
\nonumber \\
&&
\;\;\;\;\;\;\;\;\;\;\; + \;\;
\triangle m_{\mu e} (1-\cos\theta) \, J_0 \, T \, [G_1(y_0) + y_0 G_0(y_0)] \nonumber \\
&&
\;\;\;\;\;\;\;\;\;\;\; + \;\;
\triangle m_{\mu e} E_{\nu_\mu} (1-\cos\theta) \, J_1 \, G_0(y_0) \nonumber \\
&&
\;\;\;\;\;\;\;\;\;\;\; + \;\;
\left.
\triangle m_{\mu e}^2 J_2 \, G_0(y_0)
\right\}~.
\label{eq:I1}
\end{eqnarray}
The functions $\triangle$, $A$, $B$, $C$, $f_\gamma(x)$, as well as the integral functionals $G_n(y_0; \eta')$, are defined in Ref.~\cite{Mezzacappa:1993gx}, see Eqs.~(14)--(20). The latter are related to the Fermi integrals, which depend on the degeneracy parameter, $\eta'$, which is related to $\eta=\mu_e/T$, and defined as follows, 
\begin{equation}
\eta' = \eta - \frac{\left(E_{\nu_\mu} - \mu_\mu\right) - \left(E_{\nu_e} - \mu_e\right)}{T}~,
\end{equation}
in contrast to Eq.~(22) in Ref.~\cite{Mezzacappa:1993gx}, since electrons and muons can have rather different Fermi energies under SN conditions. Furthermore, the argument, $y_0$, of the Fermi-integrals, $G_n(y_0; \eta')$, also has an explicit dependence on the electron--muon rest-mass energy difference as follows,
\begin{eqnarray}
y_0 &=& \frac{1}{T}
\left\{
-\frac{1}{2} \left[
E_{\nu_\mu} - E_{\nu_e} - \left(\frac{\triangle m_{\mu e}}{E_{\nu_\mu}(1-\cos\theta)}\right)
\right]
\right. \\
&&
\qquad\qquad
\left.
+
\frac{\triangle}{2}
\left[
1
+
\frac{2\,(m_e\,c^2)^2}{E_{\nu_\mu}E_{\nu_e}(1-\cos\theta)}
+
\frac{2 \, \triangle m_{\mu e}}{E_{\nu_\mu}E_{\nu_e}(1-\cos\theta)}
+
\left(\frac{\triangle m_{\mu e}}{E_{\nu_\mu}E_{\nu_e}(1-\cos\theta)}\right)^2
\right]^{1/2}
\right\}~.
\nonumber
\end{eqnarray}
The additional phase-space terms, $J_0$, $J_1$ and $J_2$, are given by the following expressions,
\begin{eqnarray}
J_0 &=& E_{\nu_\mu}^3 + E_{\nu_\mu}^2 E_{\nu_e} \, (2 + \cos\theta) - E_{\nu_\mu} E_{\nu_e}^2 \,(2+\cos\theta) - E_{\nu_e}^3 ~, \\
J_1 &=& E_{\nu_\mu}^3 - E_{\nu_\mu}^2 E_{\nu_e} \, \cos\theta + E_{\nu_\mu} E_{\nu_e}^2 (\cos^2\theta - 2) + E_{\nu_e}^3 \, \cos\theta ~, \\
J_2 &=& E_{\nu_\mu}^2 \, \cos\theta - \frac{1}{2} E_{\nu_\mu} E_{\nu_e} (3+\cos\theta) + E_{\nu_e} ^2 \, \cos\theta~.
\end{eqnarray}

Note also, in order to eliminate the azimuthal dependence of the scattering kernels, the same 32-point Gauss quadrature numerical integration is performed as in the case of neutrino--lepton scattering. Note further that the relation of detailed balance holds here as well for the transition amplitudes of the lepton-flavor exchange processes. However, due to the presence of two different leptonic chemical potentials, the phase-space distributions for electrons and muons give rise to an additional contribution to the relation of detailed balance for the scattering kernels, as follows,
\begin{eqnarray}
&& 
\mathcal{R}^{\rm in}_{\rm LFE, \nu_\mu}(E_{\nu_\mu},E_{\nu_e},\vartheta,\vartheta')  = \mathcal{R}^{\rm out}_{\rm LFE, \nu_\mu}(E_{\nu_\mu},E_{\nu_e},\vartheta,\vartheta')  \exp\left\{-\frac{E_{\nu_\mu}-E_{\nu_e}+\mu_e-\mu_\mu}{T}\right\}~.
\label{eq:detailed_balance_leptonic_1}
\end{eqnarray}
\end{widetext}

Note that, since the transition amplitudes for the processes involving $e^+$ and $\mu^+$ are the same as for the processes involving $e^-$ and $\mu^-$, the scattering kernels are given by the same expression $I_1$~\eqref{eq:I1}, with the replacement of the chemical potentials, $\mu_{e/\mu}\rightarrow -\mu_{e/\mu}$, and $E_{\nu_{e/\mu}}\rightarrow E_{\bar\nu_{e/\mu}}$.

For the implementation of the LFE processes, (3a) and (3b) in Table~\ref{tab:muonic-weak}, in the collision integral, the scattering kernels are computed on the fly as part of the inelastic scattering module of {\tt BOLTZTRAN}. However, contrary to neutrino--lepton scattering, here the initial- and final-state neutrinos belong to different flavors. A new module for this class of inelastic processes had to be introduced according to \eqref{eq:leptonic_a_dfdt}, $\mathcal{R}_{\rm LFE, \nu}$ for (anti)muon- and (anti)electron-neutrinos.

\subsubsection{{\bf Lepton flavor conversion (LFC)}}
\label{sec:B3b}
For LFC reactions (4a) and (4b) of Table~\ref{tab:muonic-weak}, the collision integral of the Boltzmann equation takes the same form as \eqref{eq:leptonic_a_dfdt}, though changing initial- and final-state neutrino distributions respectively. Also in- and out-scattering kernels have the same algebraic structure as for the lepton flavor exchange processes \eqref{eq:Rin_leptonic_a} and \eqref{eq:Rout_leptonic_a}, i.e. converting (positron)electron into (anti)muon and vize versa. However, the matrix elements for LFC reactions are different.

In the following, the process, $\bar\nu_{e} + e^- \leftrightarrows \mu^- + \bar\nu_{\mu}$, will be discussed as an example, for which the matrix element can be directly read off from \eqref{eq:M_nms_flavor_change}, replacing the $\nu_\mu$ spinor with that of $\bar\nu_e$ neutrino and the $\nu_e$ spinor with that of $\bar\nu_\mu$ neutrino. Then, the transition amplitude takes the following form, 
\begin{equation}
\sum_{\rm s} \left\vert\mathcal{M}\right\vert^2_{\bar\nu_e + e^- \rightarrow \mu^- + \bar\nu_\mu}
=
64\,G_{\rm F}^2 \left(p_{\bar\nu_\mu}'\cdot p_{e}\right) \left(p_{\mu}' \cdot p_{\bar\nu_e}\right)~,
\label{eq:M_nms_flavor_vonc}
\end{equation}
such that the out-scattering kernel becomes,
\begin{eqnarray}
&&
\mathcal{R}^{\rm out}_{\rm LFC, \bar\nu_e}(E_{\bar\nu_\mu},E_{\bar\nu_e},\cos\theta)
=
4\frac{G_{\rm F}^2}{2\pi^2} \frac{1}{E_{\bar\nu_\mu}E_{\bar\nu_e}} 
I_2(E_{\bar\nu_\mu},E_{\bar\nu_e},\cos\theta)~,\nonumber \\
&&
\end{eqnarray}
with the same remaining phase-space integral, $I_2(E_{\bar\nu_\mu},E_{\bar\nu_e},\cos\theta)$, as for the case of neutrino--muon scattering discussed above (see also Eq.~(12) in Ref.~\cite{Mezzacappa:1993gx}), with the replacements $E_{\nu_\mu}\rightarrow -E_{\bar\nu_e}$ and $E_{\nu_e}\rightarrow -E_{\bar\nu_\mu}$, as well as the inclusion of the muon-electron rest-mass energy scale $\triangle m_{\mu e}$. Then, applying the same nomenclature as in Ref.~\cite{Mezzacappa:1993gx} and as in \eqref{eq:I1}, the resulting additional terms, $J_0$, $J_1$, and $J_2$, can be computed straightforwardly with the aforementioned replacements. Note that the scattering kernels for the processes involving $e^+$ and $\mu^+$ are obtained by the replacement  of the chemical potentials as follows, $\mu_{e/\mu}\rightarrow -\mu_{e/\mu}$.

Since in- and out-scattering LFC kernels have the same algebraic structure as in- and out-scattering LFE kernels, respectively, the reverse LFC processes are related through detailed balance in the same way the LFE kernels are \eqref{eq:detailed_balance_leptonic_1}. Hence, it is convenient to define the total lepton flavor exchange/conversion scattering kernel,
\begin{equation}
\mathcal{R}^{\rm out}_{\nu} = \mathcal{R}^{\rm out}_{\rm LFE, \nu} + \mathcal{R}^{\rm out}_{\rm LFC, \nu}~.
\end{equation}

Note that LFE and LFC reactions change the abundance of muons and electrons. Their contributions have to be taken into account by modifying the evolution equations~\eqref{eq:dfvedt} and \eqref{eq:dfvebdt} as follows,

\begin{widetext}

\begin{eqnarray}
&&
\frac{\partial Y_\mu}{\partial t} 
=
\left.\frac{\partial Y_\mu}{\partial t}\right\vert_{\rm CC}
 \\
&&
-
\left.\frac{2\pi m_{\rm B}}{\left(hc\right)^3}
\right[
\int dE_{\nu_\mu} dE_{\nu_\mu}^2 d\left(\cos\vartheta\right)
\left.\frac{\partial F_{\nu_\mu}}{\partial t}(E_{\nu_\mu},\vartheta)\right\vert_{\rm LFE+LFC}
-
\left.
\left.
\int dE_{\bar\nu_\mu} dE_{\bar\nu_\mu}^2 d\left(\cos\vartheta\right)\frac{\partial F_{\bar\nu_\mu}}{\partial t}(E_{\bar\nu_\mu},\vartheta)
\right\vert_{\rm LFE+LFC}
\right]~.
\nonumber
\end{eqnarray}
Similarly, the $Y_e$ evolution equation has to be modified, as well,
\begin{eqnarray}
&&
\frac{\partial Y_e}{\partial t} 
=
\left.\frac{\partial Y_e}{\partial t}\right\vert_{\rm CC}
\\
&&
-
\left.\frac{2\pi m_{\rm B}}{\left(hc\right)^3}
\right[
\int dE_{\nu_e} dE_{\nu_e}^2 d\left(\cos\vartheta\right)
\left.\frac{\partial F_{\nu_e}}{\partial t}(E_{\nu_e},\vartheta)\right
\vert_{\rm LFE+LFC}
-
\left.
\left.
\int dE_{\bar\nu_e} dE_{\bar\nu_e}^2 d\left(\cos\vartheta\right)
\frac{\partial F_{\bar\nu_e}}{\partial t}(E_{\bar\nu_e},\vartheta)\right
\vert_{\rm LFE+LFC}
\right]~.
\nonumber
\end{eqnarray}

\end{widetext}

\bibliography{references}

\end{document}